\newcommand{\secref}[1]{Sec.~\ref{#1}}
\newcommand{\figref}[1]{Fig.~\ref{#1}}
\newcommand{\tabref}[1]{Table~\ref{#1}}
\newcommand{\equref}[1]{Eq.~(\ref{#1})}

\documentclass[letterpaper,journal]{IEEEtran}
\usepackage{amssymb}
\usepackage{amsmath}
\usepackage{graphicx}
\usepackage{multirow}
\usepackage{algorithm}
\usepackage{algorithmicx}
\usepackage{algpseudocode}
\usepackage{epstopdf}
\usepackage{array}
\usepackage{xcolor}
\usepackage{setspace}
\usepackage{enumitem}
\usepackage{pdfpages}
\usepackage{booktabs}
\usepackage{makecell}
\usepackage{tabularx}
\usepackage{float}
\usepackage{balance}
\usepackage{hyperref}
\definecolor{mycustompurple}{RGB}{128,0,128}
\hypersetup{
	colorlinks=true,   
	linkcolor=teal,     
	citecolor=magenta,   
	filecolor=mycustompurple,   
	urlcolor=red    
}
\makeatletter

\newcommand{\Rmnum}[1]{\expandafter\@slowromancap\romannumeral #1@}
\newtheorem{remark}{Remark}
\makeatother
\begin{document}
	\title{Cognitive Link-Flexible FTN-OTFS Design for High-Mobility LEO Satellite Communications}

	\author{Chaorong Zhang, \ 
		Benjamin K. Ng, \ 
		Hui Xu,\ 
		Yue Liu, \ 
		Chan-Tong Lam,\
		Ke Wang,
		and Arumugam Nallanathan
		\thanks{
        \textsl{(Corresponding author: Benjamin K. Ng.)}
    	}}

	\maketitle

	\begin{abstract}
	Low Earth orbit (LEO) satellite links are challenged by pronounced signal-to-noise ratio (SNR) variation and severe Doppler shifts. 
	Although mobility robustness is provided by orthogonal time frequency space (OTFS) modulation, the rate-reliability tradeoff of faster-than-Nyquist (FTN) signaling is constrained by fixed packing. 
	In this paper, an SNR-aware flexible FTN-OTFS framework is proposed, in which the packing factor is adapted to changing link conditions under a prescribed reliability requirement. 
	Link-state perception and packing decisions are supported by estimated SNR and offline reliability thresholds, with constant decision complexity achieved through a fixed-mode lookup table. 
	Elevation-dependent propagation, fractional Doppler, and FTN-induced interference are incorporated into the signal model, while colored noise is accommodated by covariance-aware linear minimum mean-square error detection. 
	Throughput and bit error rate are analytically characterized, and energy efficiency and peak-to-average power ratio are evaluated. 
	Simulation results show that conservative packing preserves reliability under unfavorable conditions, while denser signaling improves throughput as the link strengthens. 
	These findings highlight the potential of cognitive FTN adaptation for efficient future LEO wireless communications.
	\end{abstract}


\section{Introduction}
\label{sec:intro}

\subsection{Low Earth Orbit Satellite}
Low Earth orbit (LEO) satellite networks extend broadband coverage beyond terrestrial infrastructure.
Compared with medium Earth orbit (MEO) and geostationary Earth orbit (GEO) satellites, their lower orbital altitude reduces propagation delay \cite{ref1,ref2}.
These advantages support communication services that require wide-area reach and responsive links.
The same orbital configuration, however, causes the satellite-user geometry to change rapidly throughout a visible pass.
The resulting motion produces time variation in propagation delay, Doppler, and link budget \cite{ref3,ref4}.
\par
As the satellite moves from the horizon toward the zenith and then returns to the horizon, the elevation-dependent slant distance, clutter loss, and atmospheric path length alter the received signal-to-noise ratio (SNR).
Consequently, the LEO link margin is not stationary over a pass.
Relative motion also makes the channel doubly selective, so the physical-layer waveform must preserve reliable detection under delay and Doppler dispersion.
Against this background, the waveform must first accommodate high mobility and then use favorable link intervals without compromising the reliability requirement.
The high-mobility aspect of this design problem motivates the use of orthogonal time frequency space modulation.
\par

\subsection{Orthogonal Time Frequency Space}
To address the mobility component of the LEO waveform design, orthogonal time frequency space (OTFS) modulation maps information symbols to the delay-Doppler (DD) domain and spreads each symbol across the time-frequency grid \cite{ref5}.
This two-dimensional mapping exposes the delay and Doppler structure of a doubly selective channel instead of treating its rapid time variation only in the time-frequency domain.
The resulting DD-domain interaction supports diversity combining across channel paths, including operation under fractional Doppler \cite{ref6}.
This delay-Doppler representation therefore provides a structured basis for receiver processing in a high-mobility link.
\par
Building on this structure, recent LEO-OTFS studies have developed sparse DD-domain channel estimation and pilot designs for high-dynamic links \cite{ref7,ref8}.
These developments motivate the use of OTFS as the modulation basis in this work because reliable channel acquisition precedes any adaptation to the evolving LEO link margin.
Despite these advantages, standard OTFS retains the Nyquist time-spacing constraint.
It improves waveform robustness to high mobility but does not by itself convert favorable link intervals into denser signaling.
This remaining rate dimension motivates the integration of faster-than-Nyquist signaling with OTFS.
\par

\subsection{Faster-than-Nyquist}
To exploit the time-domain signaling margin left by Nyquist-spaced OTFS, faster-than-Nyquist (FTN) signaling transmits pulses at the interval $T_{\mathrm{F}}=\alpha T_0$, where $T_0$ is the Nyquist interval and $0<\alpha\leq 1$ is the packing factor.
Decreasing $\alpha$ shortens the pulse interval and increases the raw signaling rate.
The denser pulse placement also strengthens pulse correlation, FTN-induced inter-symbol interference (ISI), and matched-filter noise correlation \cite{ref9,ref10,ref11}.
The resulting rate-reliability tradeoff depends on whether the available SNR and receiver processing can support the additional interference while satisfying the reliability requirement.
\par
In an LEO pass, this tradeoff becomes time dependent because the horizon-to-zenith geometry changes the available SNR and link margin \cite{ref12}.
With a fixed packing factor, aggressive packing can fail to meet the reliability requirement in low-SNR intervals, whereas conservative packing leaves usable signaling margin unexploited in favorable intervals.
The same packing factor therefore need not be suitable for every phase of a pass.
Taken together, these observations motivate a flexible FTN (FFTN) controller that selects the packing factor according to the estimated SNR.
The controller applies tighter packing only when receiver processing can accommodate the corresponding FTN-induced ISI under the reliability requirement.
This SNR-aware choice of $\alpha$ leads directly to examining how existing LEO-OTFS, FTN-OTFS, adaptive FTN, and adaptive LEO transmission schemes define their adaptation variables.
From a cognitive-communications perspective, the proposed adaptation follows a lightweight perception-reasoning-decision cycle.
The estimated SNR characterizes the link state, while offline bit error rate (BER) thresholds guide packing-factor selection and FTN waveform reconfiguration without online learning or iterative optimization.
\par

\subsection{Related Works and Motivation}
Recent OTFS studies have addressed channel acquisition and pilot-aided receiver design for high-mobility channels.
Lee \textit{et al.} developed a single-tone-based estimator for delay, integer and fractional Doppler, and channel gains \cite{ref7}, while Wang \textit{et al.} proposed time-frequency (TF)-domain pilot-aided channel estimation and equalization for fast time-varying OTFS channels \cite{ref8}.
Li \textit{et al.} adapted the modulation and coding configuration of an OTFS-based LEO link according to link quality \cite{ref13}.
Channel tracking based on LEO orbital characteristics has also been developed to reduce the pilot burden associated with frequent channel-state information (CSI) acquisition \cite{ref14}.
These methods do not adjust the FTN symbol interval or the resulting ISI level.
\par
FTN has also been integrated with OTFS through transmitter precoding, power allocation, and iterative detection.
Hong \textit{et al.} diagonalized FTN-induced interference and colored noise through eigenvalue-decomposition precoding and optimized frame-level power allocation \cite{ref15}.
Wang \textit{et al.} extended nonorthogonal packing to the time and frequency dimensions and optimized DD-domain precoding for capacity \cite{ref16}.
Receiver processing, reconfigurable intelligent surfaces-assisted transmission, and joint sensing-communication waveform design have also been studied under prescribed FTN configurations \cite{ref11,ref17,ref18}.
Beyond OTFS, Tong \textit{et al.} developed adaptive FTN transmit precoding for rapidly fading channels \cite{ref19}.
Its adaptive variable is the transmit precoder rather than the packing factor.
\par
These studies address complementary aspects of high-mobility and FTN transmission, but they do not directly translate the perceived LEO link condition into a reliability-constrained FTN packing decision.
The proposed FFTN scheme addresses this gap through a lightweight cognitive adaptation mechanism.
The estimated SNR characterizes the current link state, while offline BER thresholds guide the packing-factor decision over a finite mode set.
This design controls FTN-induced nonorthogonality without changing the modulation order or solving an online precoder optimization problem.
\tabref{tab:work_comparison} summarizes this distinction.
Taken together, this positioning connects the LEO-specific link variation to a reliability-constrained choice of the FTN packing factor.
\par

\begin{table*}[t]
\caption{Comparison With Representative Adaptive and FTN-OTFS Schemes}
\label{tab:work_comparison}
\centering
\small
\setlength{\tabcolsep}{3.2pt}
\renewcommand{\arraystretch}{1.12}
\begin{tabular}{p{2.2cm}p{2.3cm}p{3.2cm}p{2.8cm}p{1.3cm}p{2cm}p{2.6cm}}
\hline
\textbf{Scheme} & \textbf{Scenario} & \textbf{Adapted/optimized variable} & \textbf{Primary objective} & \textbf{FTN ISI} & \textbf{LEO variation} & \textbf{Link information} \\
\hline
Hong \textit{et al.} \cite{ref15} & Doubly selective & Precoder and power allocation & Mutual information & Yes & No & Statistical channel model \\
Wang \textit{et al.} \cite{ref16} & High mobility & DD-domain precoder and power allocation & Capacity & Yes & No & Channel matrix \\
Xu \textit{et al.} \cite{ref11} & V2X & Detector damping & BER and detection convergence & Yes & No & Receiver observations \\
Tong \textit{et al.} \cite{ref19} & Rapid fading & Transmit precoder & Doppler robustness and BER & Yes & No & CSI feedback \\
Li \textit{et al.} \cite{ref13} & LEO & Modulation and coding & Link adaptation & No & Yes & Link quality \\
Proposed scheme & LEO & FTN packing factor $\alpha$ & Throughput under a BER requirement & Yes & Yes & Estimated SNR, CSI, and offline BER thresholds \\
\hline
\end{tabular}
\end{table*}

\subsection{Contributions}
Overall, the main contributions of this work are summarized as follows:
	\begin{itemize}
		\item A LEO FTN-OTFS signal model is formulated by jointly incorporating elevation-dependent propagation, fractional Doppler, FTN pulse overlap, and colored matched-filter noise.
		Representative low-order tapped-delay-line (TDL) abstractions are adopted to keep the full-matrix numerical evaluation tractable while retaining characteristic delay-power and line-of-sight (LOS)/non-line-of-sight (NLOS) conditions.
	\item A reliability-constrained FFTN strategy selects $\alpha$ from a finite mode set according to the estimated link SNR.
	The rule restricts aggressive packing when its BER requirement is not satisfied and applies tighter packing when the available SNR margin permits it.
	\item Offline SNR thresholds construct the packing regions, and a lookup table implements the online decision.
	Imperfect link knowledge is separated into SNR-estimation uncertainty for mode selection and channel-estimation error for covariance-aware linear minimum mean-square error (LMMSE) detection.
        \item Effective throughput, energy efficiency (EE), analytical BER approximation, peak-to-average power ratio (PAPR), omplementary cumulative distribution function (CCDF), and computational requirements are characterized.
        Simulations with fixed FTN references evaluate the throughput-reliability behavior under representative LEO propagation profiles and estimation uncertainty.
\end{itemize}

\subsection{Organization and Notation}
\textit{Organization:} The remainder of this paper is organized as follows.
\secref{sec:system_model} presents the LEO FTN-OTFS system model, representative low-order channel abstractions, and the imperfect CSI model.
\secref{sec:reliability} formulates the reliability-constrained packing rule and its lookup-table implementation.
\secref{sec:performance} analyzes effective throughput, spectral efficiency, EE, analytical BER approximation, PAPR, and computational complexity.
Numerical results and discussions are provided in \secref{sec:simulation}, followed by conclusions in \secref{sec:conclusion}.
\par
\textit{Notation:}
The sets of complex numbers and integers are denoted by $\mathbb{C}$ and $\mathbb{Z}$, respectively, and $\mathbb{C}^{m\times n}$ denotes the space of $m\times n$ complex-valued matrices.
The $i$th entry of a vector $\mathbf{a}$ and the $(i,j)$th entry of a matrix $\mathbf{A}$ are denoted by $[\mathbf{a}]_i$ and $[\mathbf{A}]_{i,j}$, respectively.
The operators $(\cdot)^{T}$, $(\cdot)^{H}$, $(\cdot)^{*}$, and $(\cdot)^{-1}$ denote transpose, Hermitian transpose, complex conjugation, and matrix inversion, respectively.
The superscript $(\cdot)^{\star}$ identifies an optimizing solution, whereas $\widehat{(\cdot)}$ denotes an estimated quantity.
The notation $|\cdot|$ represents the absolute value of a scalar or the cardinality of a set, while $\|\cdot\|_{2}$ and $\|\cdot\|_{F}$ denote the Euclidean and Frobenius norms, respectively.
The operators $\mathbb{E}\{\cdot\}$, $\Pr\{\cdot\}$, and $\mathrm{Tr}(\cdot)$ denote statistical expectation, probability, and matrix trace, respectively.
The operator $\mathrm{vec}(\cdot)$ denotes column-wise vectorization, $\otimes$ denotes the Kronecker product, and $\odot$ denotes the Hadamard product.
The symbol $\mathbf{I}_{n}$ denotes the $n\times n$ identity matrix.
The distributions $\mathcal{N}(\mu,\sigma^{2})$ and $\mathcal{CN}(\mu,\sigma^{2})$ denote real Gaussian and circularly symmetric complex Gaussian distributions with mean $\mu$ and variance $\sigma^{2}$, respectively.
The symbols $\lfloor\cdot\rfloor$ and $\operatorname{round}(\cdot)$ denote the floor and nearest-integer operations, while $\inf$ and $\arg\min$ denote the infimum and minimizing argument, respectively.
The symbols $\delta(\cdot)$ and $\delta[\cdot]$ denote the Dirac delta function and the discrete Kronecker delta, respectively.
$j=\sqrt{-1}$ denotes the imaginary unit, and $\mathcal{O}(\cdot)$ denotes asymptotic computational complexity.
Finally, $\triangleq$ indicates a definition, and $\sim$ means ``distributed as.''

\begin{figure}
	\centering
	\includegraphics[width=\columnwidth]{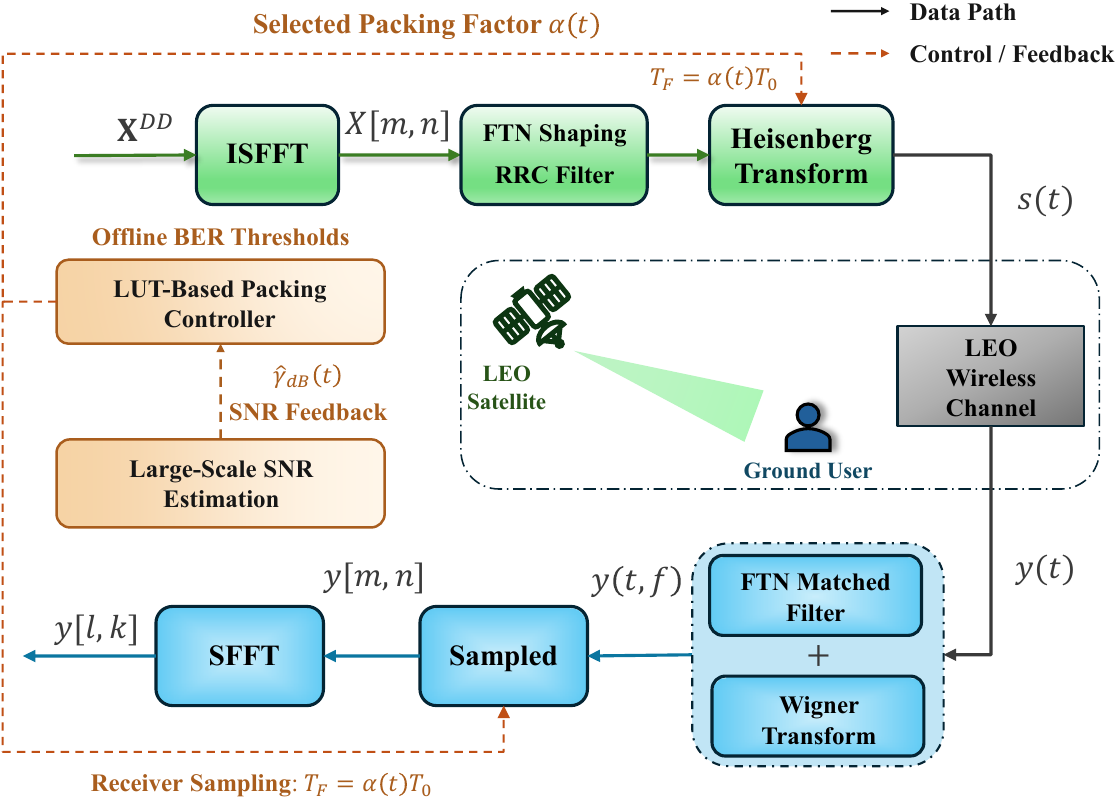}
	\caption{System model and signal processing.}
	\label{fig:system_model}
\end{figure}

\section{System Model}
\label{sec:system_model}
	\subsection{Transmitter Design and FTN Signaling}
	We first investigate a downlink LEO satellite communication single-input single-output (SISO) system employing standard FTN-aided OTFS modulation.
	The DD plane is discretized into a grid of size $M \times N$, where $M$ and $N$ denote the number of delay and Doppler bins, respectively.
	Let $\mathbf{X}^{\mathrm{DD}} \in \mathbb{C}^{M \times N}$ represent the transmitted symbol matrix, where $x[l,k]$ is a symbol drawn from the selected modulation constellation at delay index $l$ and Doppler index $k$.
	The DD-domain symbols are independent, zero-mean, and normalized to unit average energy.
	The physical energy assigned to each modulation symbol before FTN packing is denoted by $E_s$.
	The DD index sets are $\mathcal{M}=\{0,1,\ldots,M-1\}$ and $\mathcal{N}=\{0,1,\ldots,N-1\}$, respectively, with $l,m\in\mathcal{M}$ and $k,n\in\mathcal{N}$.
	\par
	The modulation process begins with the inverse symplectic fast Fourier transform (ISFFT), mapping the DD domain symbols to the TF domain as
	\begin{align}\label{eq:ISFFT}
		X[m,n] = \frac{1}{\sqrt{MN}}\sum_{l=0}^{M-1}\sum_{k=0}^{N-1} x[l,k] e^{j2\pi \left( \frac{ml}{M} - \frac{nk}{N} \right)},
	\end{align}
	where the index ranges in \equref{eq:ISFFT} follow $m\in\mathcal{M}$ and $n\in\mathcal{N}$.
	\par
\begin{figure}
		\centering
		\includegraphics[width=8.5cm,height=8.5cm]{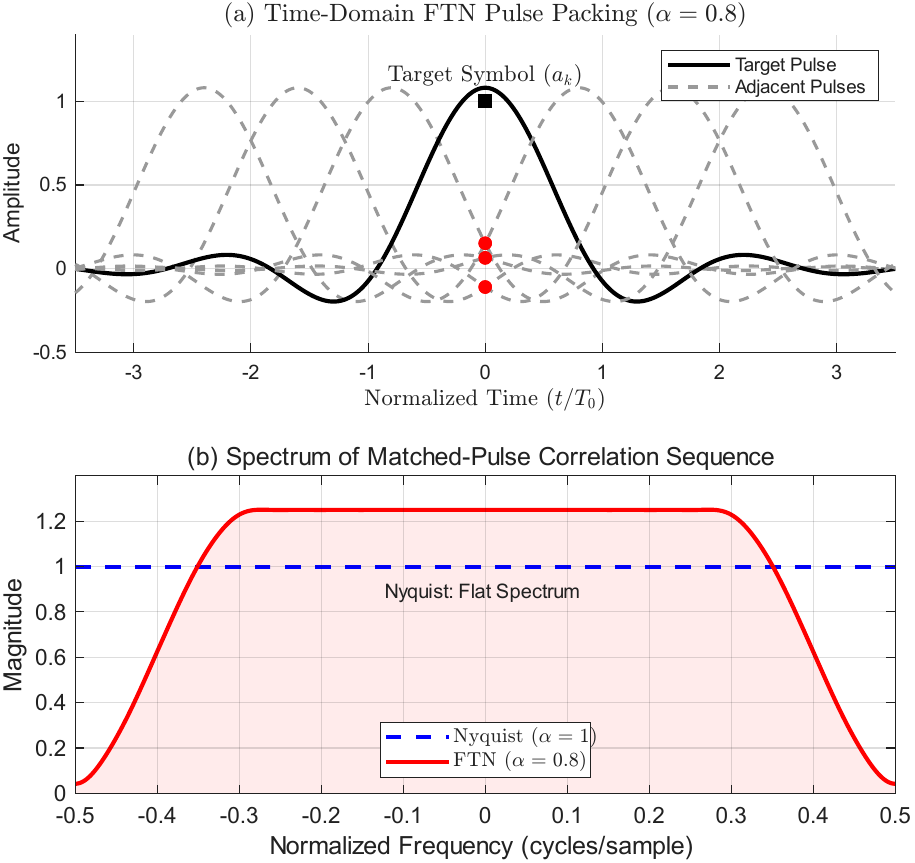}\\
		\caption{Illustration of (a) time-domain pulse packing and (b) the corresponding matched-pulse correlation spectrum under FTN signaling.}
		\label{fig:FTN}
	\end{figure}

		To enhance spectral efficiency, time-slot FTN signaling uses a packing factor $\alpha\in(0,1]$.
		The Nyquist slot duration and the packed slot duration are defined as $T_0=1/\Delta f$ and $T_{\mathrm{F}}=\alpha T_0$, respectively.
		Consequently, the Nyquist and FTN frame durations are $T_{\mathrm{fr,N}}=NT_0$ and $T_{\mathrm{fr,F}}=NT_{\mathrm{F}}=N\alpha T_0$, while $B=M\Delta f$ denotes the nominal grid bandwidth used for noise integration and spectral-efficiency normalization.
	The Heisenberg transform and transmit pulse $g_{\mathrm{tx}}(t)$ generate the continuous-time waveform
	\begin{align} \label{eq:Tx_Signal}
		s(t)
		&=\sqrt{E_s}
		\sum_{m=0}^{M-1}\sum_{n=0}^{N-1}
		X[m,n]g_{\mathrm{tx}}(t-nT_{\mathrm{F}})
		\nonumber\\
		&\quad \times
		e^{j2\pi m\Delta f(t-nT_{\mathrm{F}})}.
	\end{align}
	The pulse-shaping filter is a root-raised-cosine (RRC) pulse with roll-off factor $\beta$ and Nyquist interval $T_0$:
		\begin{align} \label{eq:RRC_Filter}
			g_{\mathrm{tx}}(t) = \frac{1}{\sqrt{T_0}}\frac{\sin\left(\pi \frac{t}{T_0} (1-\beta)\right) + 4\beta \frac{t}{T_0} \cos\left(\pi \frac{t}{T_0} (1+\beta)\right)}{\pi \frac{t}{T_0} \left[ 1 - \left(4\beta \frac{t}{T_0}\right)^2 \right]}.
		\end{align}
		The factor $1/\sqrt{T_0}$ gives $\int_{-\infty}^{\infty}|g_{\mathrm{tx}}(t)|^2dt=1$, which is consistent with $p(0)=1$ in the matched-filter noise model.
	The influence of the packing factor $\alpha$ is illustrated in \figref{fig:FTN}.
	In \figref{fig:FTN}(a), reducing $\alpha$ decreases the spacing between adjacent transmit pulses and increases their temporal overlap.
		The nonzero adjacent-pulse values at the nominal sampling instant illustrate pulse superposition, whereas the exact matched-filter coupling coefficients are determined by the pulse-correlation function $p(\tau)$ defined in \equref{eq:pulse_correlation}.
	Panel (b) does not show the power spectral density of the transmitted waveform.
	It shows the normalized magnitude response of the sampled matched-pulse correlation sequence $g_\ell(\alpha)=p(\ell T_{\mathrm{F}})$.
	Its discrete-frequency response is $G_\alpha(e^{j2\pi f})=\sum_{\ell}g_\ell(\alpha)e^{-j2\pi f\ell}$.
		Under ideal Nyquist sampling, $g_\ell(1)=\delta[\ell]$, which gives a flat response.
		For $\alpha<1$, the off-zero correlation coefficients become nonzero and produce a nonflat equivalent-channel response.
		The same correlation sequence forms the Toeplitz matrix $\mathbf{G}_{\mathrm{time}}$ in \equref{eq:G_time} and characterizes the time-domain noise correlation retained by the detector model.
\par

\subsection{LEO Wireless Environments}
The forward link channel between the LEO satellite and the ground user equipment (UE) is characterized by a combination of large-scale path loss and small-scale multipath fading.
Unlike terrestrial models that often assume static channel gains, our model explicitly accounts for the dynamic link geometry and Doppler shifts unique to LEO constellations.
\par

\subsubsection{Large-Scale Path Loss Modeling}
The forward link budget is primarily governed by the large-scale path loss $L$ (in dB), which aggregates geometric spreading, terrestrial blockage, and atmospheric impairments.
Following ITU-R P.618 \cite{ref20}, the total path loss is expressed as
\begin{align} \label{eq:PL_total}
	L = L_b + L_g + L_s,
\end{align}
where $L_b$, $L_g$, and $L_s$ denote the basic path loss, atmospheric gas attenuation, and scintillation loss, respectively.
Let $G_{\mathrm{tx,lin}}=10^{G_{\mathrm{tx,dBi}}/10}$ and $G_{\mathrm{rx,lin}}=10^{G_{\mathrm{rx,dBi}}/10}$ denote the linear transmit and receive antenna gains, respectively.
A comparable decomposition of the LEO forward-link loss has also been used in prior satellite channel modeling \cite{ref21}.
\par

The basic path loss $L_b$ accounts for the deterministic free-space propagation and stochastic environmental shadowing.
It is composed of the free-space path loss (FSPL), shadow fading (SF), and clutter loss (CL):
\begin{align} \label{eq:PL_basic}
	L_b = \mathrm{FSPL}(d, f_c) + \mathrm{SF} + \mathrm{CL}(\theta_E, f_c).
\end{align}
Here, $\mathrm{SF} \sim \mathcal{N}(0, \sigma_{\mathrm{SF}}^2)$ represents log-normal shadow fading caused by local obstacles.
The FSPL is determined by the carrier frequency $f_c$ (in GHz) and the instantaneous slant distance $d$ (in meters) as follows:
\begin{align} \label{eq:FSPL}
	\mathrm{FSPL}(d, f_c) = 32.45 + 20 \log_{10}(f_c) + 20 \log_{10}(d).
\end{align}
Given the orbital geometry of LEO constellations, the slant distance $d$ varies substantially with the time-dependent elevation angle $\theta_E$.
Based on the Earth-centered geometric model with Earth radius $R_E \approx 6371$ km and satellite altitude $h_0$, $d$ is derived as:
\begin{align} \label{eq:SlantDist}
	d = \sqrt{R_{E}^2 \sin^2(\theta_E) + h_{0}^2 + 2h_0R_{E}} - R_{E} \sin(\theta_E).
\end{align}

To accurately capture terrestrial propagation characteristics, particularly in urban areas where buildings obstruct signals, we model CL as a function of both frequency and elevation angle:
\begin{align} \label{eq:CL}
	\mathrm{CL}(\theta_E, f_c) = A_{cl} + B_{cl} \log_{10}(f_c) + C\left( 1-\sin \left( \theta_E \right) \right),
\end{align}
where $A_{cl}$ and $B_{cl}$ are frequency-dependent coefficients, and $C=20$ is an empirical factor for typical urban scenarios.
The term $(1-\sin(\theta_E))$ explicitly models the elevation dependence, reflecting that signals arriving from the zenith (high $\theta_E$) experience minimal clutter attenuation.
\par

Furthermore, the signal traverses atmospheric layers, incurring frequency-dependent absorption.
Following ITU-R P.676 \cite{ref22}, the atmospheric gas attenuation $L_g$ is modeled as
\begin{align} \label{eq:gas_attenuation}
	L_g = \frac{A_{\mathrm{zenith}}(f_c)}{\sin(\theta_E)},
\end{align}
where $A_{\mathrm{zenith}}$ denotes the zenith attenuation.
The $1/\sin(\theta_E)$ factor accounts for the increased effective atmospheric path length at lower elevation angles.
Finally, scintillation loss $L_s$, induced by rapid fluctuations in the refractive index due to ionospheric ($f_c < 6$ GHz) or tropospheric ($f_c > 6$ GHz) irregularities \cite{ref20}, is incorporated.
Note that rain attenuation is omitted in this study under the clear-sky assumption.
\par

	\begin{table*}[t!]
		\caption{Representative Low-Order TDL Parameters Used for LEO Channel Simulation}
		\label{tab:channel_params}
		\centering
		\footnotesize
		\renewcommand{\arraystretch}{1.3} 
		\setlength{\tabcolsep}{5pt}
			\begin{tabular}{|c|c|p{6cm}|p{4cm}|}
				\hline
				\textbf{Model} & \textbf{Scenario} & \textbf{Representative Small-Scale Parameters} & \textbf{Large-Scale Parameters} \\
				\hline
				\multirow{2}{*}{\textbf{TDL-A}} & Urban, NLOS & Delays: [0, 110, 285] ns & $A_{cl}=15, B_{cl}=5$ \\
				& $\theta_E = 20^\circ$ & Power: [0, $-4.7$, $-6.5$] dB, first-tap $K_{\mathrm{dB}}=-\infty$ & $\sigma_{\mathrm{SF}}=6$ dB \\
				\hline
				\multirow{2}{*}{\textbf{TDL-B}} & Urban, NLOS & Delays: [0, 105, 275] ns & $A_{cl}=15, B_{cl}=5$ \\
				& $\theta_E = 30^\circ$ & Power: [0, $-3.8$, $-5.2$] dB, first-tap $K_{\mathrm{dB}}=-\infty$ & $\sigma_{\mathrm{SF}}=6$ dB \\
				\hline
				\multirow{2}{*}{\textbf{TDL-C}} & Rural, NLOS & Delays: [0, 260, 830] ns & $A_{cl}=5, B_{cl}=2$ \\
				& $\theta_E = 30^\circ$ & Power: [0, $-3.5$, $-5.8$] dB, first-tap $K_{\mathrm{dB}}=-\infty$ & $\sigma_{\mathrm{SF}}=6$ dB \\
				\hline
				\multirow{2}{*}{\textbf{TDL-D}} & Rural, LOS & Delays: [0, 290, 895] ns & $A_{cl}=5, B_{cl}=2$ \\
				& $\theta_E = 60^\circ$ & Power: [0, $-4.2$, $-6.1$] dB, first-tap $K_{\mathrm{dB}}=13.3$ dB & $\sigma_{\mathrm{SF}}=2$ dB \\
				\hline
				\multirow{2}{*}{\textbf{TDL-E}} & Open, LOS & Delays: [0, 150, 350] ns & $A_{cl}=0, B_{cl}=0$ \\
				& $\theta_E = 85^\circ$ & Power: [0, $-8.0$, $-12.0$] dB, first-tap $K_{\mathrm{dB}}=22.0$ dB & $\sigma_{\mathrm{SF}}=2$ dB \\
				\hline
			\end{tabular}%
	\end{table*}

	\subsubsection{Small-Scale DD Domain Fading}
	Superimposed on the large-scale path loss is the small-scale multipath fading.
		To describe representative non-terrestrial propagation conditions, the large-scale geometry follows 3GPP TR 38.811, while the small-scale delay-power structures are informed by the TDL profiles in 3GPP TR 38.901 \cite{ref23,ref24}.
		A three-tap a three-tap non-terrestrial-network TDL model and LEO link-budget construction has also been used in prior high-mobility OTFS modeling \cite{ref25}.
		The five parameter sets in \tabref{tab:channel_params} are representative low-order abstractions adopted to limit the computational burden of full $MN\times MN$ matrix evaluation while retaining characteristic delay-power and LOS/NLOS conditions; they do not reproduce the complete standardized profiles.
		Each physical delay $\tau_p$ is retained explicitly in the matched-filter cross-ambiguity channel model.
	For low elevation angles or dense urban environments where the direct LOS is obstructed, the reduced TDL-A/B/C profiles represent NLOS Rayleigh fading.
	For high elevation angles or open rural environments, the reduced TDL-D/E profiles represent LOS-dominant Rician fading.
	The representative parameters are summarized in \tabref{tab:channel_params}.
	\par
	The channel impulse response in the DD domain with $P$ dominant taps is given by
	\begin{align}\label{eq:DD_channel_new}
		h(\tau,\nu)= \sum_{p=0}^{P-1} h_p\,\delta(\tau-\tau_p)\,\delta(\nu-\nu_p),
	\end{align}
	where $\tau_p$ and $\nu_p$ denote the delay and Doppler shift of the $p$-th path.
	For a tabulated first-tap factor $K_{\mathrm{dB}}$, define $K_{\mathrm{lin}}=10^{K_{\mathrm{dB}}/10}$.
	The NLOS profiles use $K_{p,\mathrm{lin}}=0$ for every path.
	For TDL-D and TDL-E, only the first path uses the tabulated Rician factor, while the remaining paths use $K_{p,\mathrm{lin}}=0$.
	The complex path coefficient is then modeled as
	\begin{align}\label{eq:Rician_coeff}
		{h_p = \sqrt{P_p} \left( \sqrt{\frac{K_{p,\mathrm{lin}}}{K_{p,\mathrm{lin}}+1}} e^{j\varphi_p} + \sqrt{\frac{1}{K_{p,\mathrm{lin}}+1}} h_{\mathrm{sc},p} \right).}
	\end{align}
	Here, $P_p$ is the normalized power of the $p$th tap with $\sum_pP_p=1$, $\varphi_p$ is the deterministic LOS phase, and $h_{\mathrm{sc},p}\sim\mathcal{CN}(0,1)$ is the diffuse component.
	\par
	For reporting the resolved DD indices, define $\Delta\tau=1/(M\Delta f)$ and map the physical delay $\tau_p$ to its nearest DD-grid location according to
	\begin{align}\label{eq:delay_doppler_param_new}
		l_p
		&=\operatorname{round}\!\left(\frac{\tau_p}{\Delta\tau}\right),
		\quad \widetilde{\tau}_p=l_p\Delta\tau,\nonumber\\
		k_p
		&=\left\lfloor\nu_pNT_{\mathrm{F}}\right\rfloor,
		\quad \kappa_p=\nu_pNT_{\mathrm{F}}-k_p\in[0,1),\nonumber\\
		\nu_p
		&=\frac{k_p+\kappa_p}{NT_{\mathrm{F}}}.
	\end{align}
	Here, $l_p\in\mathbb{Z}$ and $k_p\in\mathbb{Z}$ are the reported delay and integer Doppler indices, respectively, and $\widetilde{\tau}_p$ is the nearest DD-grid location.
	These indices describe the grid support but do not replace the physical $\tau_p$ and $\nu_p$ retained in the continuous matched-filter channel matrix; $\kappa_p$ denotes the fractional Doppler offset.
	The inverse Fourier relation between Doppler and time gives the equivalent time-varying impulse response
	\begin{align}\label{eq:time_channel_new}
		h(t,\tau)=\sum_{p=0}^{P-1} h_p\,e^{j2\pi \nu_p (t-\tau_p)}\,\delta(\tau-\tau_p).
	\end{align}

	\subsection{Received Signals and OTFS Demodulation}
	The received continuous-time baseband signal is
	        \begin{align}\label{eq:rx_time_new}
		        y(t)
		        &=\sqrt{G_{\mathrm{tx,lin}}G_{\mathrm{rx,lin}}10^{-L/10}}\nonumber\\
		        &\quad\times\sum_{p=0}^{P-1} h_p\,e^{j2\pi \nu_p (t-\tau_p)}\,s(t-\tau_p)+w(t),
	        \end{align}
        where $L$ is the path loss in decibels defined in \equref{eq:PL_total}, and $w(t)$ is zero-mean complex additive white Gaussian noise (AWGN) satisfying $\mathbb{E}\{w(t)w^*(t')\}=\sigma^2\delta(t-t')$.
        With unit-energy receive pulses, $\sigma^2$ is also the matched-filter noise variance per sample.
\par
		Under the Wigner-transform convention in \equref{eq:Wigner_new}, the matched receive pulse is $g_{\mathrm{rx}}(t)=g_{\mathrm{tx}}(t)$ because the receive pulse is already conjugated in the integral kernel.
	The Wigner transform converts $y(t)$ into the TF domain as
	\begin{align}\label{eq:Wigner_new}
		y(t,f)=\int g_{\mathrm{rx}}^{*}(t'-t)\,y(t')\,e^{-j2\pi f(t'-t)}\,dt'.
	\end{align}
	Sampling at $(t,f)=(nT_{\mathrm{F}},m\Delta f)$ yields discrete TF samples
	\begin{align}\label{eq:sample_new}
		{y[m,n]=y(t,f)\big|_{t=nT_{\mathrm{F}},\,f=m\Delta f}.}
	\end{align}
	Finally, the symplectic fast Fourier transform (SFFT) provides the DD-domain observations:
	\begin{align}\label{eq:SFFT_new}
		y[l,k]=\frac{1}{\sqrt{MN}}\sum_{m=0}^{M-1}\sum_{n=0}^{N-1}
		y[m,n]\,e^{-j2\pi\left(\frac{ml}{M}-\frac{nk}{N}\right)}.
	\end{align}
	We further define the DD-domain resolutions as
	\begin{align}\label{eq:resolutions_new}
		{\Delta\tau=\frac{1}{M\Delta f},\qquad \Delta\nu=\frac{1}{NT_{\mathrm{F}}}.}
	\end{align}
		For a path with parameters $(\tau_p,\nu_p)$, the reported delay index, integer Doppler index, and fractional Doppler offset are $l_p$, $k_p$, and $\kappa_p$, respectively, as defined in \equref{eq:delay_doppler_param_new}.
	\par
	Let $\mathbf{x}^{\mathrm{DD}}\triangleq \mathrm{vec}(\mathbf{X}^{\mathrm{DD}})$ and $\mathbf{y}^{\mathrm{DD}}\triangleq \mathrm{vec}(\mathbf{Y}^{\mathrm{DD}})$ denote the vectorized DD-domain transmit and receive blocks, respectively.
	Column-wise vectorization is adopted throughout, so the TF sample at $(m,n)$ has index $q=m+nM$, while the input sample at $(m',n')$ has index $q'=m'+n'M$.
	Define $\mathbf{x}_{\mathrm{TF}}\triangleq\mathrm{vec}(\mathbf{X}_{\mathrm{TF}})$ and $\mathbf{y}_{\mathrm{TF}}\triangleq\mathrm{vec}(\mathbf{Y}_{\mathrm{TF}})$ using the same ordering.
	For $Q\in\{M,N\}$, let $\mathbf{F}_Q$ be the normalized discrete Fourier transform (DFT) matrix with entries $[\mathbf{F}_Q]_{a,b}=Q^{-1/2}e^{-j2\pi ab/Q}$.
	The matrix forms of the ISFFT and SFFT are, respectively,
	\begin{align}\label{eq:ISFFT_SFFT_matrices}
		\mathbf{U}_{\mathrm{I}}
		&=\mathbf{F}_N\otimes\mathbf{F}_M^H,\nonumber\\
		\mathbf{U}_{\mathrm{S}}
		&=\mathbf{U}_{\mathrm{I}}^H
		=\mathbf{F}_N^H\otimes\mathbf{F}_M.
	\end{align}
	Both $\mathbf{U}_{\mathrm{I}}$ and $\mathbf{U}_{\mathrm{S}}$ are unitary $MN\times MN$ matrices.
	Consistent with \equref{eq:SFFT_new}, the corresponding vector transformations are
	\begin{align}\label{eq:TF_DD_vector_transforms}
		\mathbf{x}_{\mathrm{TF}}
		&=\mathbf{U}_{\mathrm{I}}\mathbf{x}^{\mathrm{DD}},\nonumber\\
		\mathbf{y}^{\mathrm{DD}}
		&=\mathbf{U}_{\mathrm{S}}\mathbf{y}_{\mathrm{TF}}.
	\end{align}
	To construct the effective channel without introducing an implicit sampling time, we first formulate the pulse-shaped matched-filter channel in the TF domain.
	The cross-ambiguity function is defined as
	\begin{align}\label{eq:ambiguity_def}
		\mathcal{A}_{g}(\tau,\nu)\triangleq
		\int_{-\infty}^{\infty} g_{\mathrm{tx}}(t)\,g_{\mathrm{rx}}^{*}(t-\tau)\,e^{-j2\pi \nu t}\,dt,
	\end{align}
	which accounts for FTN-induced nonorthogonality through the sampling interval $T_{\mathrm{F}}=\alpha T_0$.
	For $q=m+nM$ and $q'=m'+n'M$, direct evaluation of the matched-filter samples in \equref{eq:sample_new} gives the TF-domain channel entry
	\begin{align}
		\label{eq:H_TF_entry}
		\left[\mathbf{H}_{\mathrm{TF}}\right]_{q,q'}
		&=
		\sqrt{E_sG_{\mathrm{tx,lin}}G_{\mathrm{rx,lin}}10^{-L/10}}
		\sum_{p=0}^{P-1}
		h_p e^{j\phi_{p,m,n,m',n'}}
		\nonumber\\
		&\quad\times
		\mathcal{A}_{g}\!\left(
		(n-n')T_{\mathrm{F}}-\tau_p,\,
		(m-m')\Delta f-\nu_p
		\right),
	\end{align}
	where the phase term is
	\begin{align}\label{eq:H_TF_phase}
		\phi_{p,m,n,m',n'}
		&=2\pi\!\left[\nu_p n'T_{\mathrm{F}}
		-m\Delta f\bigl(\tau_p+(n'-n)T_{\mathrm{F}}\bigr)\right].
	\end{align}
	Thus, $\mathbf{H}_{\mathrm{TF}}\in\mathbb{C}^{MN\times MN}$ incorporates the transmit-energy scaling, physical path Ddelays and Doppler shifts, large-scale link gain, and FTN pulse overlap.
	The on-grid quantities in \equref{eq:delay_doppler_param_new} remain useful for reporting the resolved DD parameters, but they do not replace $\tau_p$ and $\nu_p$ in the matched-filter relation \equref{eq:H_TF_entry}.
	The TF-domain and DD-domain input--output relations are therefore
	\begin{align}\label{eq:TF_DD_IO_new}
		\mathbf{y}_{\mathrm{TF}}
		&=\mathbf{H}_{\mathrm{TF}}\mathbf{x}_{\mathrm{TF}}
		+\mathbf{z}_{\mathrm{TF}},\nonumber\\
		\mathbf{y}^{\mathrm{DD}}
		&=\mathbf{H}_{\mathrm{eff}}\mathbf{x}^{\mathrm{DD}}
		+\mathbf{z}^{\mathrm{DD}},
	\end{align}
	where the effective DD-domain channel and noise are
	\begin{align}\label{eq:DD_IO_new}
		\mathbf{H}_{\mathrm{eff}}
		&=\mathbf{U}_{\mathrm{S}}\mathbf{H}_{\mathrm{TF}}
		\mathbf{U}_{\mathrm{I}},\nonumber\\
		\mathbf{z}^{\mathrm{DD}}
		&=\mathbf{U}_{\mathrm{S}}\mathbf{z}_{\mathrm{TF}}.
	\end{align}
	Because $\mathbf{H}_{\mathrm{TF}}$, $\mathbf{U}_{\mathrm{I}}$, and $\mathbf{U}_{\mathrm{S}}$ are all $MN\times MN$, $\mathbf{H}_{\mathrm{eff}}\in\mathbb{C}^{MN\times MN}$.
	The receiver operates on this full matrix and does not assume block diagonalization.
    \par
    Nonorthogonal FTN pulse shaping and matched filtering produce correlated discrete-time noise samples.
    The TF-domain vector $\mathbf{z}_{\mathrm{TF}}\in\mathbb{C}^{MN\times 1}$ follows a zero-mean circularly symmetric complex Gaussian distribution with covariance
    \begin{align}\label{eq:colored_noise_new}
        {\mathbb{E}\{\mathbf{z}_{\mathrm{TF}}\mathbf{z}_{\mathrm{TF}}^{H}\}
        = \sigma^2 \mathbf{G}_{\mathrm{TF}}.}
    \end{align}
    Under the adopted frequency-orthogonality approximation, inter-subcarrier noise correlation is neglected, while the time-domain correlation induced by FTN pulse packing is retained along the $N$ time indices.
    With the ordering $q=m+nM$, this approximation yields the following $MN\times MN$ TF-domain correlation matrix:
    \begin{align}\label{eq:G_TF}
        {\mathbf{G}_{\mathrm{TF}}
        =\mathbf{G}_{\mathrm{time}}\otimes\mathbf{I}_M.}
    \end{align}
    Here, $\mathbf{I}_M$ is the $M\times M$ identity matrix, and $\mathbf{G}_{\mathrm{time}}\in\mathbb{C}^{N\times N}$ is the Hermitian Toeplitz autocorrelation matrix determined by the pulse-shaping filter and the time-domain FTN packing factor $\alpha$.
    Define the matched-pulse correlation function as
    \begin{align}\label{eq:pulse_correlation}
        {p(\tau)=\int_{-\infty}^{\infty}g_{\mathrm{tx}}(t)
        g_{\mathrm{tx}}^{*}(t-\tau)\,dt.}
    \end{align}
    The Toeplitz correlation matrix is then given in \equref{eq:G_time}.
    \begin{figure*}[!t]
    \centering
        \begin{align}     \label{eq:G_time}
			\mathbf{G}_{\mathrm{time}} =
        \begin{bmatrix}
            g_0 & g_{-1}  & \cdots & g_{-(N-1)} \\ g_1 & g_0  & \cdots & g_{-(N-2)} \\ g_2 & g_1  & \cdots & g_{-(N-3)} \\ \vdots & \vdots & \vdots & \ddots & \vdots \\ g_{N-1} & g_{N-2} & \cdots & g_0
        \end{bmatrix}
        =
        \begin{bmatrix}
            {p(0)} & {p(-T_{\mathrm{F}})} & {p(-2T_{\mathrm{F}})} & \cdots & {p(-(N-1)T_{\mathrm{F}})} \\ {p(T_{\mathrm{F}})} & {p(0)} & {p(-T_{\mathrm{F}})} & \cdots & {p(-(N-2)T_{\mathrm{F}})} \\ {p(2T_{\mathrm{F}})} & {p(T_{\mathrm{F}})} & {p(0)} & \cdots & {p(-(N-3)T_{\mathrm{F}})} \\ \vdots & \vdots & \vdots & \ddots & \vdots \\ {p((N-1)T_{\mathrm{F}})} & {p((N-2)T_{\mathrm{F}})} & {p((N-3)T_{\mathrm{F}})} & \cdots & {p(0)}
        \end{bmatrix}
	\end{align}
	\hrulefill
    \end{figure*}
    Applying the SFFT in \equref{eq:DD_IO_new} to the TF-domain noise gives the DD-domain covariance
    \begin{align}\label{eq:Rz_DD}
        \mathbf{R}_z
        &=\mathbb{E}\{\mathbf{z}^{\mathrm{DD}}(\mathbf{z}^{\mathrm{DD}})^H\}\nonumber\\
        &=\sigma^2\mathbf{U}_{\mathrm{S}}\mathbf{G}_{\mathrm{TF}}
        \mathbf{U}_{\mathrm{S}}^H.
    \end{align}
    \par
    The average matched-filter noise power per TF sample, denoted by $P_z$, follows from the covariance trace as
    \begin{align}\label{eq:noise_power}
        P_z
        &= \frac{1}{MN} \mathbb{E}\{ \mathbf{z}_{\mathrm{TF}}^H \mathbf{z}_{\mathrm{TF}} \} \nonumber \\
        &= \frac{\sigma^2}{MN}
        \mathrm{Tr}(\mathbf{G}_{\mathrm{time}}\otimes\mathbf{I}_M).
    \end{align}
    Because the pulse is energy normalized, $p(0)=1$ and $\mathrm{Tr}(\mathbf{G}_{\mathrm{time}})=N$.
    The Kronecker trace identity then yields
    \begin{align}\label{eq:noise_power_result}
        {P_z = \frac{\sigma^2}{MN}
        \mathrm{Tr}(\mathbf{G}_{\mathrm{time}})\mathrm{Tr}(\mathbf{I}_M)
        =\sigma^2.}
    \end{align}
    Thus, $\alpha$ changes the correlation structure but preserves the average matched-filter noise power per symbol under the adopted pulse normalization.
    Unitary transformation by $\mathbf{U}_{\mathrm{S}}$ also preserves this trace in the DD domain.
\par

	\subsection{Imperfect CSI Model}
	\label{subsec:imperfect_csi}
		In a high-dynamic LEO link, the DD-domain channel used by the detector is an estimate of the actual channel.
		To define a normalized effective-channel perturbation, let $\mathbf{Z}$ contain independent $\mathcal{CN}(0,1)$ entries and define the binary mask by $[\mathbf{M}_H]_{i,j}=1$ when $[\mathbf{H}_{\mathrm{eff}}]_{i,j}\neq0$ and $[\mathbf{M}_H]_{i,j}=0$ otherwise.
		A random error direction is formed and normalized as
	\begin{align}\label{eq:CSI_error_direction}
		{\widetilde{\mathbf{E}}_0=\mathbf{M}_H\odot\mathbf{Z},\qquad
		\mathbf{E}_0=\frac{\|\mathbf{H}_{\mathrm{eff}}\|_F}
		{\|\widetilde{\mathbf{E}}_0\|_F}\widetilde{\mathbf{E}}_0.}
	\end{align}
		For a normalized channel-estimation-error power $\delta_H^2$, the error and estimated channel are
	\begin{align}\label{eq:CSI_error_model}
		{\mathbf{E}_H=\sqrt{\delta_H^2}\,\mathbf{E}_0,\qquad
		\widehat{\mathbf{H}}_{\mathrm{eff}}
		=\mathbf{H}_{\mathrm{eff}}+\mathbf{E}_H}.
	\end{align}
		This construction gives $\|\mathbf{E}_H\|_F^2/\|\mathbf{H}_{\mathrm{eff}}\|_F^2=\delta_H^2$ for each generated realization.
		The received vector in \equref{eq:TF_DD_IO_new} is generated by the actual $\mathbf{H}_{\mathrm{eff}}$, whereas the detector is constructed from $\widehat{\mathbf{H}}_{\mathrm{eff}}$ and retains the full colored-noise covariance $\mathbf{R}_z$.
		This channel-estimation error is distinct from the SNR-estimation uncertainty used only for packing-mode selection.
		Frequent CSI acquisition also consumes pilot resources in an LEO link, which motivates this explicit separation \cite{ref14}.
\par

	\subsection{Signal Detection}
		A covariance-aware LMMSE detector is used to recover $\mathbf{x}^{\mathrm{DD}}$.
		The unit-energy symbol normalization gives $\mathbb{E}\{\mathbf{x}^{\mathrm{DD}}
		(\mathbf{x}^{\mathrm{DD}})^H\}=\mathbf{I}_{MN}$.
		The factor $\sqrt{E_s}$ is included in $\mathbf{H}_{\mathrm{TF}}$ and hence in
		$\mathbf{H}_{\mathrm{eff}}$.
		The detector constructed from the available channel estimate is
	\begin{align}\label{eq:LMMSE_new}
		\widehat{\mathbf{x}}^{\mathrm{DD}}
		=\widehat{\mathbf{W}}{\mathbf{y}^{\mathrm{DD}}},
	\end{align}
	\begin{align}\label{eq:LMMSE_new2}
		\widehat{\mathbf{W}}
		=\left(\widehat{\mathbf{H}}_{\mathrm{eff}}^H\mathbf{R}_z^{-1}
		\widehat{\mathbf{H}}_{\mathrm{eff}}+\mathbf{I}_{MN}\right)^{-1}\widehat{\mathbf{H}}_{\mathrm{eff}}^H\mathbf{R}_z^{-1}.
	\end{align}
	For perfect CSI, $\widehat{\mathbf{H}}_{\mathrm{eff}}=\mathbf{H}_{\mathrm{eff}}$.
	Under imperfect CSI, the same expression uses \equref{eq:CSI_error_model}.
	The covariance $\mathbf{R}_z$ in \equref{eq:Rz_DD} is defined in the DD domain and therefore avoids inserting the TF-domain correlation matrix directly into the detector.
\par

\section{Reliability-Constrained FFTN Signaling}
\label{sec:reliability}
	The proposed FFTN scheme controls FTN-induced ISI by selecting $\alpha$ from the ordered candidate set $\mathcal{B}=\{\alpha_1,\ldots,\alpha_K\}=\{1.0,0.9,0.8\}$.
	The packing controller uses the large-scale received SNR, which varies with the link loss $L(t)$.
	In linear units, this common link-quality variable is
	\begin{align}\label{eq:inst_SNR}
		\gamma(t)=\frac{E_{s,r}(t)}{\sigma^2}
		=\frac{E_sG_{\mathrm{tx,lin}}G_{\mathrm{rx,lin}}10^{-L(t)/10}}{\sigma^2},
	\end{align}
		where all candidate modes use the same pre-packing symbol energy $E_s$.
		For the transmit-power representation in \figref{fig:error}(a), let $P_t$ denote the Nyquist-equivalent reference transmit power in linear units and let $P_{t,\mathrm{dBm}}$ denote its decibel representation; the associated symbol energy is $E_s=P_tT_0/M$.
		The actual average radiated power of mode $\alpha$ and the corresponding reference link-budget mapping are
		\begin{align}\label{eq:link_budget_snr}
			P_{\mathrm{tx,dBm}}(\alpha)
			&=P_{t,\mathrm{dBm}}-10\log_{10}\alpha,\nonumber\\
			P_{r,\mathrm{N,dBm}}(t)
			&=P_{t,\mathrm{dBm}}+G_{\mathrm{tx,dBi}}+G_{\mathrm{rx,dBi}}-L(t),\nonumber\\
			P_{n,\mathrm{dBm}}
			&=-174+10\log_{10}(M\Delta f)+F_{\mathrm{NF}},\nonumber\\
			\gamma_{\mathrm{dB}}(t)
			&=P_{r,\mathrm{N,dBm}}(t)-P_{n,\mathrm{dBm}}.
		\end{align}
		The integrated noise power in linear units is $P_n=B\sigma^2$, where $B=M\Delta f$.
		Thus, the nominal noise-integration bandwidth $M\Delta f$, antenna gains, receiver noise figure, and large-scale propagation loss map the reference power $P_t$ to the large-scale received SNR used by the packing controller, while the physical packed-mode power follows the first line of \equref{eq:link_budget_snr}.
	Let $P_b(\gamma,\alpha_k)$ denote the BER averaged for a specified OTFS grid, modulation, representative low-order channel profile, and detector configuration.
	With reliability requirement $P_{\mathrm{th}}$, the desired packing factor is
	\begin{align}\label{eq:optimization}
		\alpha^{\star}(t)=\arg\min_{\alpha_k\in\mathcal{B}}\ \alpha_k \nonumber
		\\
		\mathrm{s.t.}\quad P_b(\gamma(t),\alpha_k)\leq P_{\mathrm{th}}.
	\end{align}
	For each candidate mode, an offline BER curve defines
	\begin{align}\label{eq:reliability_threshold}
		{\Gamma_k=\inf\left\{\gamma:P_b(\gamma,\alpha_k)\leq P_{\mathrm{th}}\right\},
		\qquad \Gamma_{k,\mathrm{dB}}=10\log_{10}\Gamma_k.}
	\end{align}
	If a mode never satisfies the requirement over the evaluated SNR range, its threshold is set to $+\infty$.
	The online controller perturbs the large-scale received SNR only for mode selection:
	\begin{align}\label{eq:SNR_error_model}
		{\widehat{\gamma}_{\mathrm{dB}}(t)
		=\gamma_{\mathrm{dB}}(t)+\Delta_{\gamma}(t),\qquad
		\Delta_{\gamma}(t)\sim\mathcal{N}(0,\sigma_\gamma^{2}).}
	\end{align}
	Based on the estimated SNR and using the thresholds $\Gamma_{1.0,\mathrm{dB}}$, $\Gamma_{0.9,\mathrm{dB}}$, and $\Gamma_{0.8,\mathrm{dB}}$, the packing rule and reliability-outage flag can be obtained as
	\begin{align}\label{eq:adjustment_rule}
		{(\alpha(t),o_{\mathrm{rel}}(t))=
		\begin{cases}
			(0.8,0), & \widehat{\gamma}_{\mathrm{dB}}(t)\geq\Gamma_{0.8,\mathrm{dB}},\\ (0.9,0), & \Gamma_{0.9,\mathrm{dB}}\leq\widehat{\gamma}_{\mathrm{dB}}(t)<\Gamma_{0.8,\mathrm{dB}},\\ (1.0,0), & \Gamma_{1.0,\mathrm{dB}}\leq\widehat{\gamma}_{\mathrm{dB}}(t)<\Gamma_{0.9,\mathrm{dB}},\\ (1.0,1), & \widehat{\gamma}_{\mathrm{dB}}(t)<\Gamma_{1.0,\mathrm{dB}}.
		\end{cases}}
	\end{align}
	The last branch selects $\alpha=1$ as the minimum-ISI fallback but explicitly records that the reliability requirement is not met.
	SNR-estimation uncertainty in \equref{eq:SNR_error_model} can change the selected mode, whereas channel-estimation error in \equref{eq:CSI_error_model} changes the LMMSE weights.

\begin{algorithm}[t]
	\caption{Reliability-Constrained FFTN Packing Selection}
	\label{alg:adaptive_ftn}
	\begin{algorithmic}[1]
		\Require Candidate set $\mathcal{B}=\{1.0,0.9,0.8\}$; thresholds $\Gamma_{1.0,\mathrm{dB}}$, $\Gamma_{0.9,\mathrm{dB}}$, and $\Gamma_{0.8,\mathrm{dB}}$; estimated SNR $\widehat{\gamma}_{\mathrm{dB}}(t)$
		\Ensure Packing factor $\alpha(t)$ and outage flag $o_{\mathrm{rel}}(t)$
		\State $o_{\mathrm{rel}}(t)\leftarrow0$
		\If{$\widehat{\gamma}_{\mathrm{dB}}(t)\geq\Gamma_{0.8,\mathrm{dB}}$}
			\State $\alpha(t)\leftarrow0.8$
		\ElsIf{$\widehat{\gamma}_{\mathrm{dB}}(t)\geq\Gamma_{0.9,\mathrm{dB}}$}
			\State $\alpha(t)\leftarrow0.9$
		\ElsIf{$\widehat{\gamma}_{\mathrm{dB}}(t)\geq\Gamma_{1.0,\mathrm{dB}}$}
			\State $\alpha(t)\leftarrow1.0$
		\Else
			\State $\alpha(t)\leftarrow1.0$, $o_{\mathrm{rel}}(t)\leftarrow1$
		\EndIf
		\State Configure $T_{\mathrm{F}}=\alpha(t)T_0$
	\end{algorithmic}
\end{algorithm}

	For a general $K$-mode set, an ordered threshold scan requires $\mathcal{O}(K)$ comparisons.
	The three-mode lookup table in this work has fixed $K$ and therefore has $\mathcal{O}(1)$ control-decision complexity.
	This statement does not include channel estimation, waveform processing, or LMMSE detection.

\vspace{-10pt}
\section{Performance Analysis}
\label{sec:performance}
This section characterizes the throughput-reliability mechanism of the proposed FFTN scheme.
Effective throughput and analytical BER approximation form the primary evaluation measures.
EE and PAPR are retained as supporting metrics.
\footnote{Unless otherwise stated, the SNR-domain BER and throughput comparisons use a common pre-packing symbol energy $E_s$ for all candidate packing factors.
This normalization isolates the effects of signaling density, FTN-induced ISI, and colored noise when $\alpha$ is varied.
The corresponding numerical results are therefore normalized SNR-domain comparisons rather than fixed-average-transmit-power comparisons.}
\par

\subsection{Throughput Analysis}
\label{sec:throughput}

\subsubsection{Raw Transmission Rate and Spectral Efficiency}
Let $M_{\mathrm{mod}}$ denote the modulation order.
The total information payload in bits per OTFS frame is given by
\begin{align}\label{eq:total_bits}
	N_{\mathrm{bits}} = MN \log_2 M_{\mathrm{mod}}.
\end{align}
In the SNR-aware FFTN-OTFS framework, $N$ packed time slots give the frame duration $T_{\mathrm{fr,F}}=N\alpha T_0$.
Consequently, the raw data rate is
\begin{align}\label{eq:raw_rate}
	{R_{\mathrm{raw}}(\alpha)
	=\frac{MN\log_2M_{\mathrm{mod}}}{N\alpha T_0}
	=\frac{M\log_2M_{\mathrm{mod}}}{\alpha T_0}.}
\end{align}
Using the nominal grid bandwidth $B=M\Delta f=M/T_0$, the raw spectral efficiency is
\begin{align}\label{eq:raw_SE}
	{\Psi_{\mathrm{raw}}(\alpha)
	=\frac{R_{\mathrm{raw}}(\alpha)}{B}
	=\frac{\log_2M_{\mathrm{mod}}}{\alpha}\quad[\mathrm{bits/s/Hz}].}
\end{align}
These raw expressions isolate the nominal $1/\alpha$ acceleration before accounting for decoding errors.
\par

\subsubsection{Effective Throughput}
In practical high-mobility LEO links, the theoretical acceleration is compromised by transmission reliability.
We define the effective throughput $\mathcal{T}_{\mathrm{eff}}$ as the net rate of successfully decoded bits.
Let $P_b(\gamma,\alpha)$ denote the BER as a function of the large-scale received SNR $\gamma$ and the interference level determined by $\alpha$.
Under an independent-bit-error approximation, the probability of at least one erroneous bit in a frame is
\begin{align}\label{eq:FER}
	P_{\mathrm{frame}}(\gamma,\alpha)
	\approx 1-\left(1-P_b(\gamma,\alpha)\right)^{N_{\mathrm{bits}}}.
\end{align}
FTN detection errors can be correlated, so \equref{eq:FER} is a throughput approximation rather than an exact frame error rate expression.
The effective throughput is the raw rate multiplied by the frame-success probability and is approximated using \equref{eq:FER} as
\begin{align}\label{eq:eff_throughput}
	\mathcal{T}_{\mathrm{eff}}(\gamma,\alpha)
	&=R_{\mathrm{raw}}(\alpha)
	\left[1-P_{\mathrm{frame}}(\gamma,\alpha)\right] \notag\\
	&\approx
	\frac{M\log_2 M_{\mathrm{mod}}}{\alpha T_0}
	\left[1-P_b(\gamma,\alpha)\right]^{N_{\mathrm{bits}}}.
\end{align}
Defining the large-scale channel power gain as
$\mathcal{H}(\theta_E)\triangleq
G_{\mathrm{tx,lin}}G_{\mathrm{rx,lin}}10^{-L(\theta_E)/10}$,
the effective throughput is approximated as
\begin{align}\label{eq:throughput_elevation}
	\mathcal{T}_{\mathrm{eff}}(\theta_E,\alpha)
	&\approx
	\frac{M\log_2 M_{\mathrm{mod}}}{\alpha T_0}
	\nonumber\\
	&\quad\times
	\left[
	1-P_b\left(
	\frac{E_s\mathcal{H}(\theta_E)}{\sigma^2},\alpha
	\right)
	\right]^{N_{\mathrm{bits}}}.
\end{align}
\par
For comparison among different packing modes, the effective throughput is normalized by the Nyquist raw rate as
\begin{align}\label{eq:nyquist_rate}
	{R_{\mathrm{N}}
	=\frac{M\log_2 M_{\mathrm{mod}}}{T_0}.}
\end{align}
The normalized effective throughput is then defined as
\begin{align}\label{eq:norm_eff_throughput}
	\overline{\mathcal{T}}_{\mathrm{eff}}(\gamma,\alpha)
	&=\frac{\mathcal{T}_{\mathrm{eff}}(\gamma,\alpha)}{R_{\mathrm{N}}}
	\nonumber\\
	&=\frac{1}{\alpha}
	\left(1-P_{\mathrm{frame}}(\gamma,\alpha)\right)
	\nonumber\\
	&\approx
	\frac{1}{\alpha}
	\left(1-P_b(\gamma,\alpha)\right)^{N_{\mathrm{bits}}}.
\end{align}
This dimensionless metric separates the rate acceleration factor $1/\alpha$ from the reliability penalty and is used for the numerical throughput comparisons.

\vspace{-10pt}
\subsection{EE}
\label{sec:EE}
The SNR-domain analysis adopts a common pre-packing energy $E_s$ for each TF/DD modulation symbol and for all candidate packing factors.
Therefore, the average radiated power depends on the pulse interval.
For the Nyquist reference with $T_{\mathrm{F}}=T_0$, the average transmit power is defined as
\begin{align}\label{eq:nyquist_tx_power}
		P_{\mathrm{N}}=\frac{ME_s}{T_0}.
\end{align}
When FTN signaling reduces the pulse interval to $T_{\mathrm{F}}=\alpha T_0$, the corresponding average transmit power is
\begin{align}\label{eq:ftn_tx_power}
		P_{\mathrm{tx}}(\alpha)
		=\frac{ME_s}{\alpha T_0}
		=\frac{P_{\mathrm{N}}}{\alpha}.
\end{align}
Hence, under the adopted common-symbol-energy normalization, tighter packing increases the signaling density and average radiated power by the same factor $1/\alpha$.
\par
Let $\xi\in(0,1]$ denote the power-amplifier (PA) drain efficiency and let $P_c$ denote the $\alpha$-independent circuit power associated with the transmitter radio-frequency (RF) and baseband components.
The total power consumption is modeled as
\begin{align}\label{eq:total_power_EE}
	P_{\mathrm{tot}}(\alpha)
	=
	\frac{P_{\mathrm{N}}}{\alpha\xi}+P_c.
\end{align}
The effective EE is defined as the number of successfully delivered information bits per unit energy,
\begin{align}\label{eq:EE_definition_new}
	\eta_{\mathrm{EE}}(\gamma,\alpha)
	=
	\frac{\mathcal{T}_{\mathrm{eff}}(\gamma,\alpha)}
	{P_{\mathrm{tot}}(\alpha)}.
\end{align}
Using the effective throughput in \equref{eq:eff_throughput}, the EE becomes
\begin{align}\label{eq:EE_final_new}
	\eta_{\mathrm{EE}}(\gamma,\alpha)
	&=
	\frac{
		\dfrac{R_{\mathrm{N}}}{\alpha}
		\left[1-P_{\mathrm{frame}}(\gamma,\alpha)\right]
	}
	{
		\dfrac{P_{\mathrm{N}}}{\alpha\xi}+P_c
	}
	\nonumber\\
	&=
	\frac{
		R_{\mathrm{N}}
		\left[1-P_{\mathrm{frame}}(\gamma,\alpha)\right]
	}
	{
		P_{\mathrm{N}}/\xi+\alpha P_c
	},
\end{align}
where
\begin{align}
	{R_{\mathrm{N}}=\frac{M\log_2M_{\mathrm{mod}}}{T_0}}
\end{align}
is the Nyquist raw transmission rate defined in \equref{eq:nyquist_rate}.
\par
\equref{eq:EE_final_new} separates the reliability effect from the circuit-energy contribution.
The explicit $1/\alpha$ acceleration in the effective throughput is accompanied by the same $1/\alpha$ increase in average radiated power under the common-symbol-energy normalization.
Therefore, tighter packing does not directly produce an RF-energy-efficiency gain.
Its EE depends on whether the reduction in circuit energy associated with the shorter transmission interval compensates for the reliability loss caused by FTN-induced interference.
The model in \equref{eq:EE_final_new} does not include channel-estimation overhead, feedback power, mode-dependent baseband processing, or PA back-off.
Accordingly, the EE results characterize the packing-dependent transmission tradeoff under the stated power model rather than the total onboard energy consumption.
\par
For a dimensionless comparison, the EE is further normalized with respect to the error-free Nyquist reference as
\begin{align}\label{eq:normalized_EE}
	\overline{\eta}_{\mathrm{EE}}(\gamma,\alpha)
	&=
	\frac{\eta_{\mathrm{EE}}(\gamma,\alpha)}{R_{\mathrm{N}}/(P_{\mathrm{N}}/\xi+P_c)}\nonumber\\
	&=
	\frac{(P_{\mathrm{N}}/\xi+P_c)\left[1-P_{\mathrm{frame}}(\gamma,\alpha)\right]}
	{P_{\mathrm{N}}/\xi+\alpha P_c}.
\end{align}
By defining the circuit-to-PA power ratio as
\begin{align}
	\rho_c = \frac{P_c}{P_{\mathrm{N}}/\xi},
\end{align}
\equref{eq:normalized_EE} can be rewritten as
\begin{align}\label{eq:normalized_EE_compact}
	\overline{\eta}_{\mathrm{EE}}(\gamma,\alpha) = \frac{1+\rho_c}{1+\alpha\rho_c}
	\left[1-P_{\mathrm{frame}}(\gamma,\alpha)\right].
\end{align}
\remark{\equref{eq:normalized_EE_compact} reveals that the EE behavior of FTN packing is governed by both transmission reliability and circuit-power consumption.
If $\rho_c=0$, the normalized EE reduces to $\overline{\eta}_{\mathrm{EE}}(\gamma,\alpha)=1-P_{\mathrm{frame}}(\gamma,\alpha)$, which indicates that tighter packing provides no intrinsic RF-energy advantage under the common-symbol-energy normalization.
When $\rho_c>0$ and the frame error probability approaches zero, the normalized EE approaches
\begin{align}
	\overline{\eta}_{\mathrm{EE}}(\gamma,\alpha)
	\rightarrow
	\frac{1+\rho_c}{1+\alpha\rho_c}.
\end{align}
For $\alpha<1$, this value exceeds unity because the shortened transmission interval reduces the circuit-energy expenditure per delivered frame.
More generally, a packing mode $\alpha$ provides a higher EE than the Nyquist mode only when
\begin{align}
	1-P_{\mathrm{frame}}(\gamma,\alpha) > \frac{1+\alpha\rho_c}{1+\rho_c}
		\left[1-P_{\mathrm{frame}}(\gamma,1)\right].
\end{align}
Thus, aggressive FTN packing is energy-efficient only when its circuit-energy reduction is not offset by the corresponding reliability loss.}

\vspace{-10pt}
\subsection{Analytical BER Approximation}
\label{sec:ber_approximation}
The covariance-aware LMMSE model also provides a streamwise BER approximation.
Under perfect CSI, let $\mathbf{e}_{\mathrm{p}}=\widehat{\mathbf{x}}^{\mathrm{DD}}-\mathbf{x}^{\mathrm{DD}}$.
Its estimation-error covariance is \cite{ref26}
\begin{align}\label{eq:lmmse_error_covariance}
	\boldsymbol{\Phi}_{\mathrm{p}}
	=\mathbb{E}\!\left[\mathbf{e}_{\mathrm{p}}
	\mathbf{e}_{\mathrm{p}}^H\right]
	=\left(\mathbf{I}_{MN}+\mathbf{H}_{\mathrm{eff}}^H
	\mathbf{R}_z^{-1}\mathbf{H}_{\mathrm{eff}}\right)^{-1}.
\end{align}
For the matched MMSE detector, the corresponding post-detection signal-to-interference-plus-noise ratio (SINR) is
\begin{align}\label{eq:perfect_csi_sinr}
	{\gamma_{\mathrm{eff},i}^{(\mathrm{p})}
	=\frac{1}{[\boldsymbol{\Phi}_{\mathrm{p}}]_{ii}}-1.}
\end{align}
The identity in \equref{eq:perfect_csi_sinr} is not applied to a mismatched detector.
Under channel-estimation error, define
\begin{align}\label{eq:mismatched_effective_matrix}
	{\mathbf{A}=\widehat{\mathbf{W}}\mathbf{H}_{\mathrm{eff}}.}
\end{align}
Because the received vector is generated by the actual channel while $\widehat{\mathbf{W}}$ is constructed from the estimated channel, the $i$th-stream SINR is evaluated directly as
\begin{align}\label{eq:imperfect_csi_sinr}
	{\gamma_{\mathrm{eff},i}^{(\mathrm{mis})}
	=\frac{|A_{ii}|^2}
	{\displaystyle\sum_{j\ne i}|A_{ij}|^2+
	[\widehat{\mathbf{W}}\mathbf{R}_z\widehat{\mathbf{W}}^H]_{ii}}.}
\end{align}
Let $\gamma_{\mathrm{eff},i}$ denote $\gamma_{\mathrm{eff},i}^{(\mathrm{p})}$ for perfect CSI and $\gamma_{\mathrm{eff},i}^{(\mathrm{mis})}$ for mismatched detection.
Under a Gaussian residual-interference approximation, the conditional BPSK BER of stream $i$ is approximated as \cite{ref27,ref28,ref29}
\begin{align}\label{eq:bpsk_ber_approximation}
	{P_{b,i}^{\mathrm{BPSK}}
	\approx Q\!\left(\sqrt{2\gamma_{\mathrm{eff},i}}\right).}
\end{align}
For square $M_{\mathrm{mod}}$-QAM, including QPSK, the corresponding approximation is
\begin{align}\label{eq:qam_ber_approximation}
	P_{b,i}^{\mathrm{QAM}}
	\approx\frac{4}{\log_2M_{\mathrm{mod}}}
	\left(1-\frac{1}{\sqrt{M_{\mathrm{mod}}}}\right) Q\!\left(\sqrt{\frac{3\gamma_{\mathrm{eff},i}}
	{M_{\mathrm{mod}}-1}}\right).
\end{align}
The frame-level analytical approximation is obtained by averaging the applicable streamwise values,
\begin{align}\label{eq:average_ber_approximation}
	{P_b(\gamma,\alpha)\approx\frac{1}{MN}
	\sum_{i=1}^{MN}P_{b,i},}
\end{align}
and then averaging over the channel realizations of the specified grid, modulation, representative low-order profile, and detector configuration.
The Gaussian residual model is an analytical approximation rather than a BER bound.
It can therefore produce analytical--simulation discrepancies when the structured FTN-induced interference is not well represented by a Gaussian variable, particularly at high SNR.
This approximation is used only for analytical BER prediction and is not used to calibrate the lookup table (LUT) thresholds. 

\vspace{-10pt}
\subsection{PAPR and CCDF Analysis}
The PAPR analysis uses the same continuous-time waveform $s(t;\alpha)$ defined in \equref{eq:Tx_Signal}; no separate waveform model is introduced.
For a LUT-selected packing factor $\alpha$, the PAPR over the FTN frame interval is \cite{ref30}
\begin{align}\label{eq:papr_definition}
	{\Xi(\alpha)=
	\frac{\displaystyle\max_{0\leq t<T_{\mathrm{fr,F}}}|s(t;\alpha)|^2}
	{\displaystyle\frac{1}{T_{\mathrm{fr,F}}}
	\int_{0}^{T_{\mathrm{fr,F}}}|s(t;\alpha)|^2\,dt}.}
\end{align}
The denominator in \equref{eq:papr_definition} is the average frame power of the packed waveform.
From \equref{eq:Tx_Signal}, the unit magnitude of the subcarrier phase term gives the following triangle-inequality bound on the peak term:
\begin{align}\label{eq:peak_power_bound}
	P_{\mathrm{peak}}(\alpha)
	&=\max_t|s(t;\alpha)|^2
	\nonumber\\
	&\leq E_s\max_t\left(
	\sum_{m=0}^{M-1}\sum_{n=0}^{N-1}
	|X[m,n]|\,|g_{\mathrm{tx}}(t-nT_{\mathrm{F}})|
	\right)^2.
\end{align}
This bound describes possible coherent pulse superposition but does not impose an unproved scaling law for the peak or average term.
For a fixed packing mode $\alpha_k$, define the conditional CCDF as $P_{\mathrm{CCDF}}(\Xi_0\mid\alpha_k)=\Pr\{\Xi(\alpha_k)>\Xi_0\}$.
Let $p_k$ be the empirical fraction of samples assigned to mode $\alpha_k$ under the specified SNR or transmit-power sampling distribution.
The aggregate CCDF follows from the law of total probability \cite{ref31}:
\begin{align}\label{eq:total_ccdf}
	{\overline{P}_{\mathrm{CCDF}}(\Xi_0)
	=\sum_{k=1}^{K}p_kP_{\mathrm{CCDF}}(\Xi_0\mid\alpha_k),
	\qquad \sum_{k=1}^{K}p_k=1.}
\end{align}
\equref{eq:total_ccdf} characterizes only the waveform PAPR distribution induced by mode occupancy.
Without a nonlinear PA model, it does not imply a particular PA back-off or energy saving.

\vspace{-10pt}
\subsection{Computational Complexity Analysis}
The ISFFT and SFFT require $\mathcal{O}(MN\log(MN))$ operations.
Let $L_f$ denote the number of retained discrete-time pulse-shaping samples processed per TF/DD symbol.
Transmit pulse shaping and matched filtering then require $\mathcal{O}(L_fMN)$ operations.
The adopted receiver directly inverts the full $MN\times MN$ matrix in \equref{eq:LMMSE_new2}; its dense implementation therefore requires $\mathcal{O}((MN)^3)$ operations and $\mathcal{O}((MN)^2)$ matrix storage.
The packing controller requires $\mathcal{O}(K)$ comparisons for a general mode set and $\mathcal{O}(1)$ comparisons for the fixed three-mode set.
These control operations do not change the cubic order of dense LMMSE detection.

\begin{table*}[t!]
	\caption{Common and Figure-Specific Simulation Parameters}
	\label{tab:sim_params}
	\centering
	\footnotesize
	\renewcommand{\arraystretch}{1.25}
	\setlength{\tabcolsep}{4.5pt}

	\begin{tabularx}{\textwidth}{
			>{\raggedright\arraybackslash}p{3.2cm}
			>{\raggedright\arraybackslash}X
			>{\centering\arraybackslash}p{2.0cm}}
		\toprule
		\textbf{Parameter} &
		\textbf{Value or Configuration} &
		\textbf{Scope} \\
		\midrule

		\multicolumn{3}{l}{\textit{\textbf{Common Parameters}}} \\
		\addlinespace[2pt]

		Orbital geometry &
		$R_E=6371$ km and $h_0=780$ km &
		\multirow{4}{*}{Common} \\

		Carrier and grid bandwidth &
		$f_c=28$ GHz, $\Delta f=15$ kHz, and $B=M\Delta f$ &
		\\

		Pulse shaping &
		RRC pulse, $\beta=0.3$, $L_{\mathrm{span}}=6$ symbols, and oversampling factor 8 &
		\\

		Atmospheric terms &
		$A_{\mathrm{zenith}}=0.22$ dB and scintillation loss $L_s=0.13$ dB &
		\\

		\midrule

		\multicolumn{3}{l}{\textit{\textbf{Figure-Specific Parameters}}} \\
		\addlinespace[2pt]

		Analytical BER comparison &
		TDL-A, $M=N=32$, BPSK, fixed $\alpha\in\{1,0.9,0.8\}$,
		and SNR-fluctuation standard deviation
		$\sigma_{\mathrm{fluc}}\in\{0,3\}$ dB &
		\figref{fig:theo} \\

		TDL-profile BER evaluation &
		TDL-A/C/D/E, $M=N=16$, QPSK, fixed FTN with $\alpha=0.9$, proposed FFTN with $\mathcal{B}=\{1,0.9,0.8\}$, and switching points of 14 and 26 dB &
		\figref{fig:TDL} \\

		Main FFTN evaluation &
		TDL-C, $M=32$, $N=16$, QPSK, $\mathcal{B}=\{1,0.9,0.8\}$, switching points of 14 and 26 dB, and $\rho_c=0.5$ &
		\figref{fig:throu} \\

		SNR-estimation uncertainty &
		TDL-B, BPSK, Nyquist-equivalent reference $P_t=10{:}5{:}50$ dBm, $G_{\mathrm{tx}}+G_{\mathrm{rx}}=50$ dB, $F_{\mathrm{NF}}=5$ dB, $\theta_E=45^\circ$; $M=N=16$ with $\sigma_\gamma\in\{0,3,5\}$ dB, and $M=N\in\{32,64\}$ with $\sigma_\gamma=3$ dB &
		\figref{fig:error}(a) \\

		Channel-estimation error &
		TDL-B, $M=N=16$, BPSK, $\alpha=0.9$, and $\delta_H^2\in\{0,0.03,0.05\}$ &
		\figref{fig:error}(b) \\

		PAPR-specific evaluation &
		$M=N=16$, 8-QAM, $\alpha\in\{1,0.95,0.9\}$, and switching points of 0 and 15 dB &
		\figref{fig:PAPR} \\

		\bottomrule
	\end{tabularx}
\end{table*}

	\begin{figure}
		\centering
		\includegraphics[width=8.5cm,height=8.3cm]{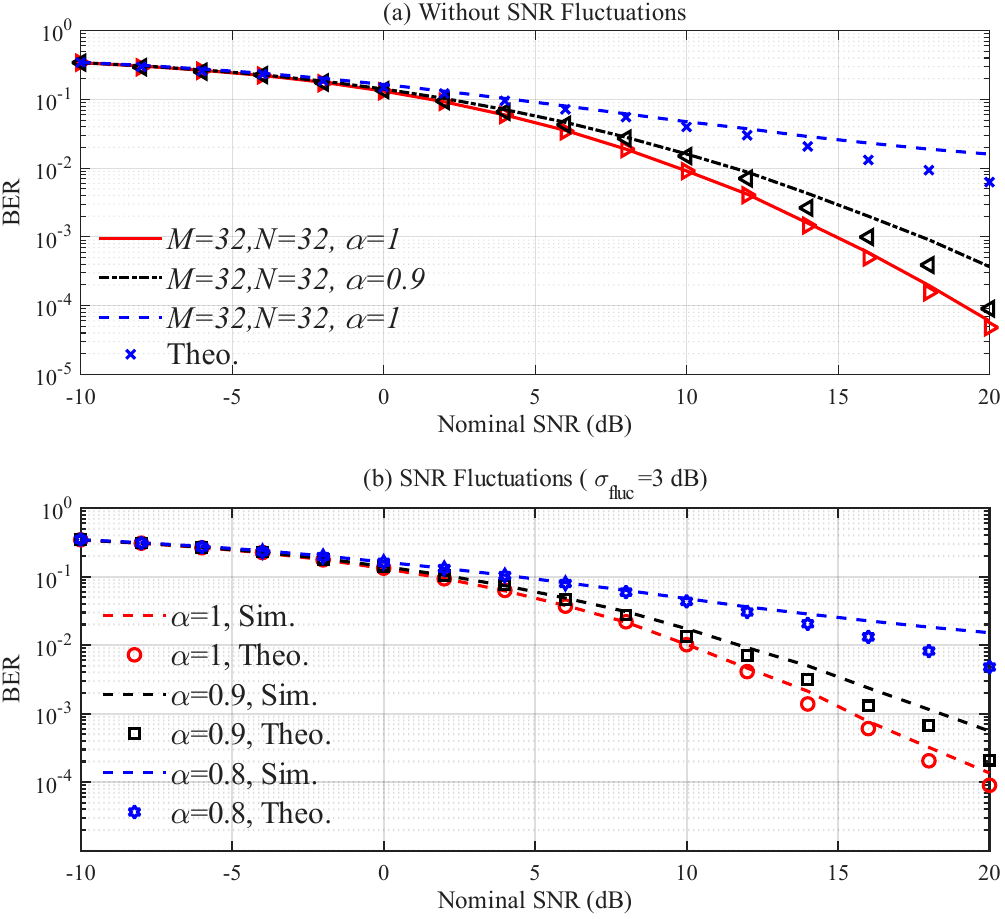}\\
		\caption{Analytical and simulated BER for fixed packing factors $\alpha\in\{1,0.9,0.8\}$ over the representative low-order TDL-A channel with $M=N=32$ and BPSK:
		(a) without SNR fluctuations and
		(b) with Gaussian SNR fluctuations having a standard deviation of $3$ dB in the dB domain.}
		\label{fig:theo}
	\end{figure}

	\begin{figure}
		\centering
		\includegraphics[width=8.5cm,height=6.3cm]{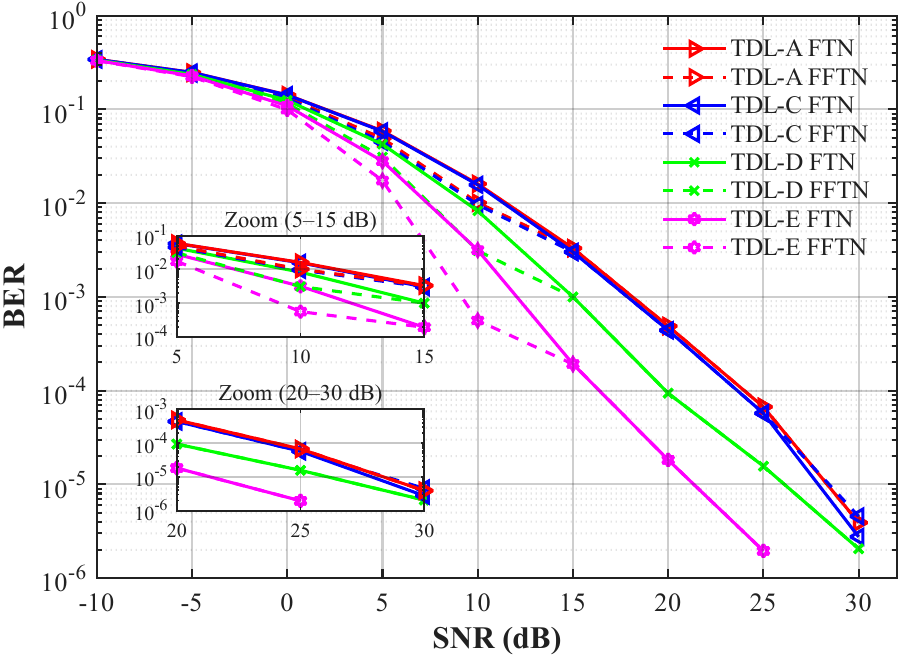}\\
		\caption{BER performance of the proposed FFTN scheme and fixed FTN references under representative low-order TDL profiles with $M=N=16$.}
		\label{fig:TDL}
	\end{figure}

\vspace{-10pt}
\section{Numerical Results}
\label{sec:simulation}
The LEO satellite altitude is $h_0=780$ km, the carrier frequency is $f_c=28$ GHz, and the subcarrier spacing is $\Delta f=15$ kHz.
Hence, $T_0=1/\Delta f\approx66.7~\mu$s and each plotted grid uses the nominal noise-integration bandwidth $B=M\Delta f$.
Regarding the FTN signaling parameters, a RRC pulse with a roll-off factor $\beta = 0.3$ is adopted, and the filter truncation length is set to $L_{\text{span}} = 6$ symbols to balance complexity and performance.
The BER and throughput evaluations use BPSK or QPSK as specified for each figure, while the independent PAPR configuration in \figref{fig:PAPR} uses 8-QAM.
The large-scale LEO geometry and loss terms follow \secref{sec:system_model}, while the small-scale fading uses the representative low-order TDL abstractions in \tabref{tab:channel_params}.
Unless otherwise specified, the BER simulations use $6\times10^{4}$ channel realizations per SNR point.
The normalized throughput and EE results are obtained from the corresponding block-error statistics of the packing modes.
The key simulation parameters are summarized in \tabref{tab:sim_params}.
\par
\begin{figure}
	\centering
	\includegraphics[width=8.5cm,height=8.7cm]{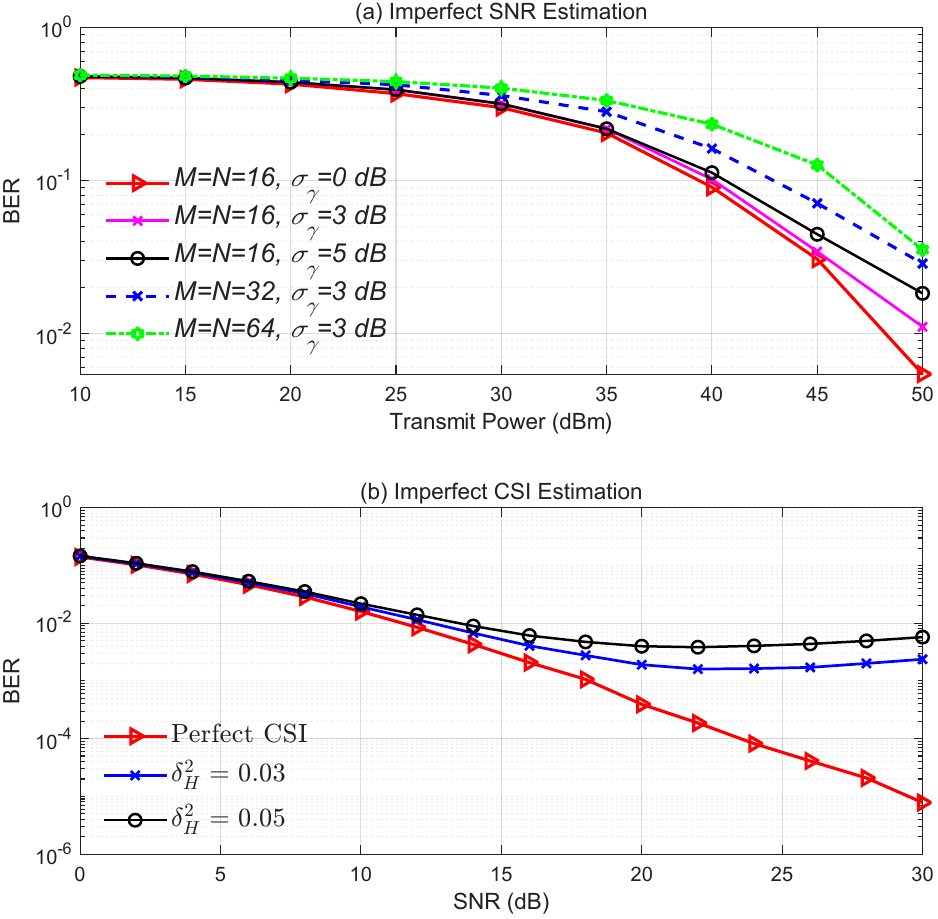}\\
	\caption{BER sensitivity under imperfect link knowledge: (a) SNR-estimation uncertainty in the FFTN packing decision for the plotted OTFS grid sizes versus the Nyquist-equivalent reference transmit power $P_t$, with the received SNR obtained from \equref{eq:link_budget_snr}, and (b) channel-estimation error for the representative low-order TDL-B channel with $M=N=16$, BPSK, and $\alpha=0.9$, where $\delta_H^2$ is the normalized channel-estimation-error power.}
	\label{fig:error}
\end{figure}

The analytical BER results in \secref{sec:ber_approximation} are compared with Monte Carlo results in \figref{fig:theo} for fixed packing modes.
Panel (a) uses the nominal SNR without additional fluctuations.
For $\alpha=1$, the analytical approximation agrees reasonably well with the simulated trend over the plotted range.
With tighter packing, the analytical and simulated curves depart more visibly at high SNR, which indicates that the Gaussian residual-interference approximation becomes less accurate for structured FTN-induced interference.
The flattening of the simulated curves over the displayed SNR interval is reported only as a finite-range numerical observation; it is neither caused by the analytical approximation nor interpreted here as a proved asymptotic error floor.
Panel (b) evaluates BER under fluctuations in the actual received SNR while keeping $\alpha$ fixed.
The actual SNR in decibels is modeled as $\gamma_{\mathrm{act,dB}}=\gamma_{\mathrm{nom,dB}}+\epsilon$, where $\epsilon\sim\mathcal{N}(0,\sigma_{\mathrm{fluc}}^2)$ and $\sigma_{\mathrm{fluc}}=3$ dB.
The horizontal axis denotes $\gamma_{\mathrm{nom,dB}}$. 
SNR-estimation uncertainty in packing selection is evaluated separately in \figref{fig:error}(a).
\par
\figref{fig:TDL} compares the BER performance across the representative low-order TDL profiles.
The TDL-E profile yields the lowest BER among the tested profiles, followed by TDL-D, whereas TDL-A yields the highest BER.
This performance ordering reflects the combined delay-power structures, LOS/NLOS conditions, Rician factors, and large-scale parameters adopted for the four representative profiles.
At low SNR, the proposed FFTN scheme achieves lower BER than the fixed FTN reference with \(\alpha=0.9\). 
In the intermediate SNR region, their BER curves coincide, consistent with the selection of \(\alpha=0.9\) between the switching thresholds of 14 and 26 dB.

	\begin{figure}
	\centering
	\includegraphics[width=8.5cm,height=9.3cm]{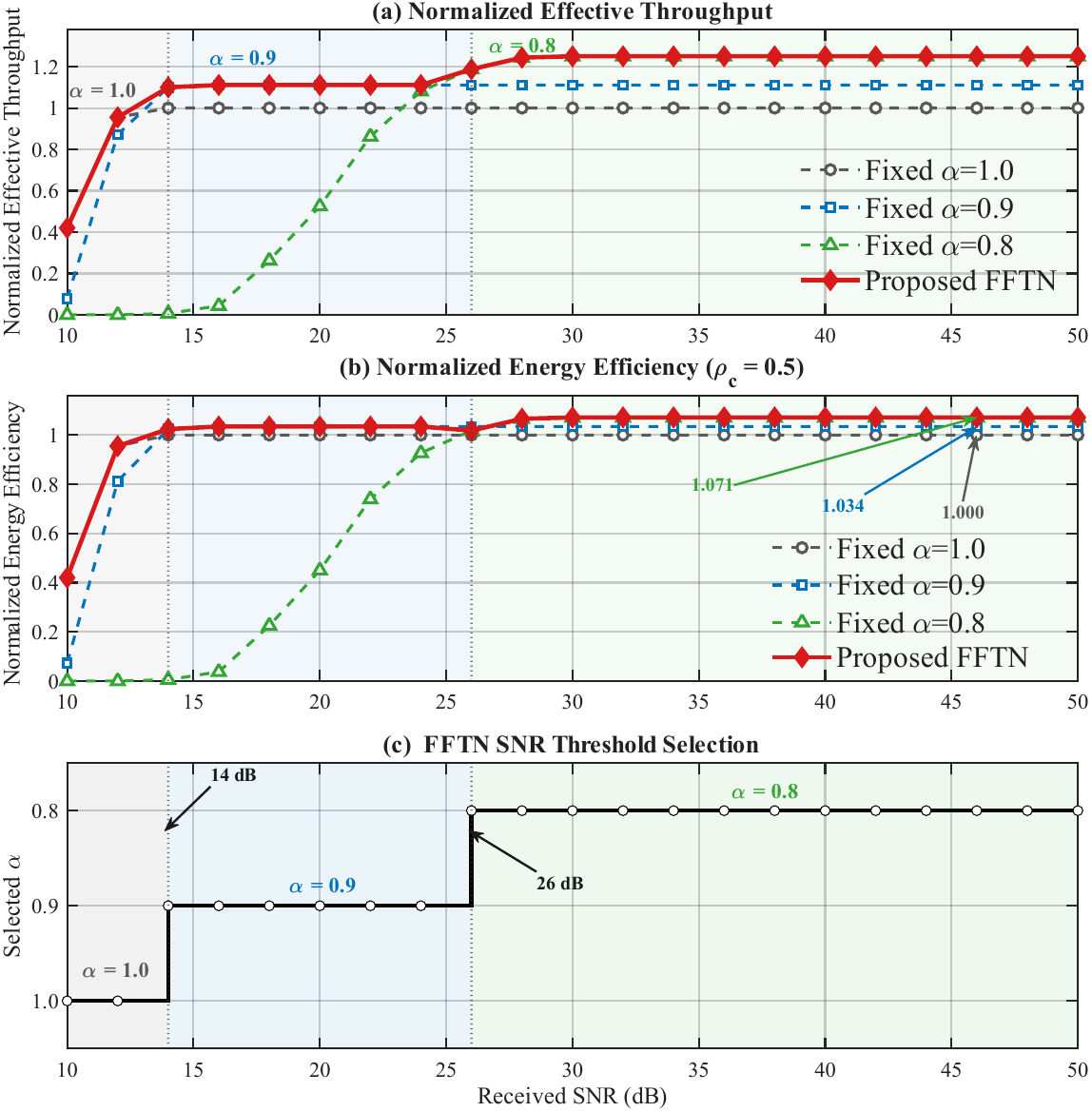}\\
	\caption{Performance of the proposed FFTN scheme and fixed packing modes with $\alpha\in\{1.0,0.9,0.8\}$: (a) normalized effective throughput, (b) normalized energy efficiency, and (c) reliability-based packing-factor selection, where $M=32$, $N=16$, $M_{\mathrm{mod}}=4$, $\rho_c=0.5$, and the adopted lookup-table switching points are 14 and 26 dB.}
	\label{fig:throu}
\end{figure}

\par
\figref{fig:throu} jointly illustrates the normalized effective throughput, normalized EE, and the corresponding packing-factor adaptation of the proposed FFTN scheme.
As shown in \figref{fig:throu}(a), the fixed modes with $\alpha\in\{1.0,0.9,0.8\}$ exhibit distinct operating regions because the rate advantage of tighter packing is realized only after the associated FTN interference becomes sufficiently manageable.
At low SNR, the aggressive $\alpha=0.8$ mode suffers from a high frame error rate and therefore provides little effective throughput, whereas the Nyquist mode preserves reliability but is limited to a normalized rate of unity.
As the SNR increases, the reliability penalty of FTN is gradually reduced, allowing $\alpha=0.9$ and then $\alpha=0.8$ to exploit their larger signaling rates.
In the high-SNR region, the normalized throughput of $\alpha=0.8$ approaches $1/\alpha=1.25$, corresponding to a 25\% increase in raw signaling rate over the Nyquist reference under the adopted common symbol-energy normalization.
A first insight is that the intermediate $\alpha=0.9$ mode is not redundant: it provides a useful transition between the robust Nyquist mode and the more aggressive $\alpha=0.8$ mode, thereby extending the SNR range over which FTN packing can be applied without incurring a severe reliability loss.
The same behavior is reflected in \figref{fig:throu}(b).
For $\rho_c=0.5$, the fixed $\alpha=0.8$ mode exhibits poor EE at low SNR because its frame-error penalty outweighs the circuit-energy reduction enabled by the shorter transmission interval, whereas its EE advantage appears only after reliable transmission is established.
The normalized EE then approaches the asymptotic values predicted by \equref{eq:normalized_EE_compact}, namely $1$, $1.034$, and $1.071$ for $\alpha=1.0$, $0.9$, and $0.8$, respectively.
This leads to a second insight: the EE improvement of tighter packing does not arise from an intrinsic reduction in RF energy, but from reducing the circuit-active duration once the reliability requirement is satisfied.
The resulting mode transitions are summarized in \figref{fig:throu}(c), where the adopted switching points change the packing factor from $\alpha=1.0$ to $\alpha=0.9$ at approximately 14 dB and then to $\alpha=0.8$ at approximately 26 dB.
For the plotted configuration, these switching points separate the operating regions in which the three modes produce different throughput and EE outcomes.
\par
\figref{fig:error} evaluates the sensitivity of the proposed FFTN scheme to SNR-estimation uncertainty and channel-estimation error.
In \figref{fig:error}(a), the horizontal axis is the Nyquist-equivalent reference transmit power $P_t$, which is mapped to the large-scale received SNR by \equref{eq:link_budget_snr}; the actual average radiated power of each packed mode scales as $1/\alpha$.
The SNR estimate used for packing selection is then perturbed according to \equref{eq:SNR_error_model}.
For $M=N=16$, increasing $\sigma_\gamma$ from $0$ to $5$ dB causes a moderate BER degradation because the estimation error can lead to packing-mode mismatch, while the physical received SNR remains unchanged.
For $\sigma_\gamma=3$ dB, the larger grids also exhibit higher BER under the tested settings.
Under the adopted $B=M\Delta f$ configuration, increasing $M$ enlarges the nominal noise-integration bandwidth and therefore increases the integrated receiver noise power at fixed reference transmit power.
The observed difference consequently reflects the combined effects of grid size and bandwidth-dependent received SNR rather than an isolated grid-size effect.
From the cognitive adaptation perspective, $\sigma_\gamma$ characterizes uncertainty in the link-perception stage and can therefore lead to an incorrect packing decision near the switching thresholds.
In \figref{fig:error}(b), imperfect CSI directly perturbs the LMMSE detector.
Increasing $\delta_H^2$ from $0.03$ to $0.05$ progressively degrades the BER, and both imperfect-CSI curves flatten over the plotted high-SNR range, whereas the perfect-CSI curve continues to decrease.
This finite-range behavior is consistent with channel-estimation mismatch becoming the dominant impairment as the thermal-noise contribution is reduced; no asymptotic error floor is inferred from the displayed range.
\par

\begin{figure}
	\centering
	\includegraphics[width=8.5cm,height=10.5cm]{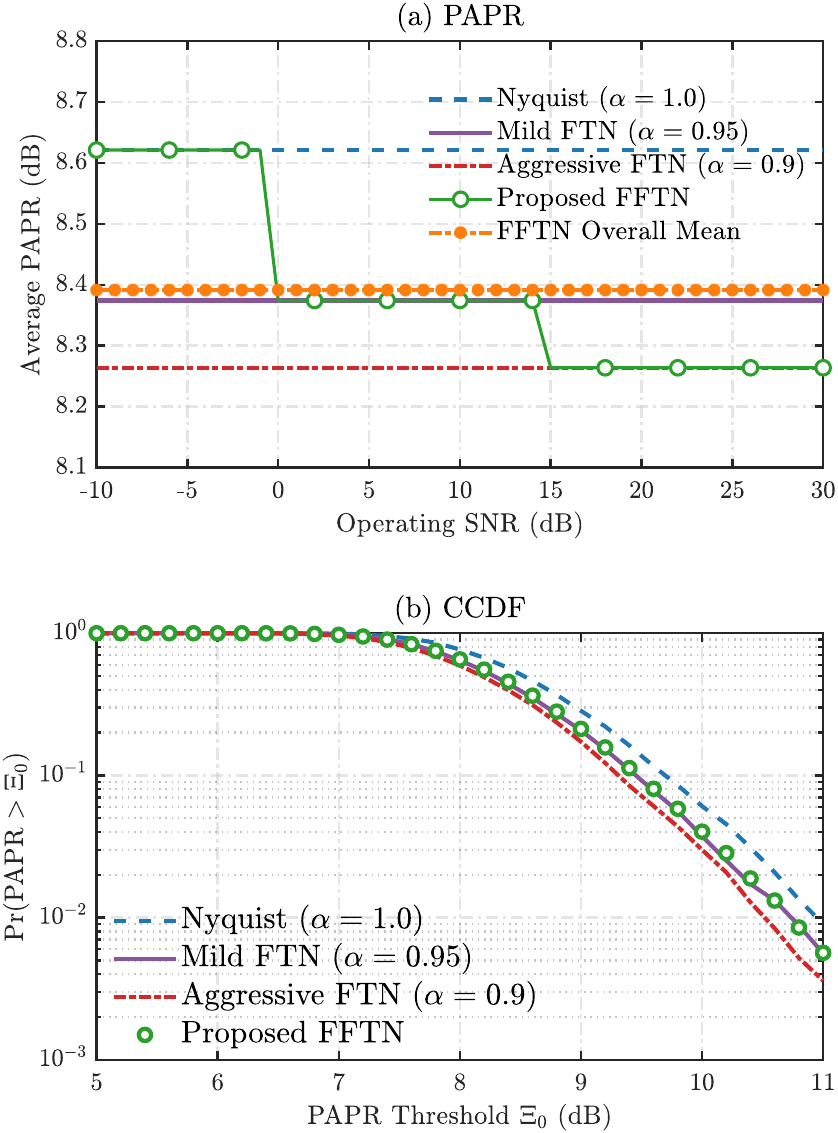}\\
	\caption{PAPR-specific evaluation with $M=N=16$, 8-QAM, packing modes $\alpha\in\{1,0.95,0.9\}$, and switching points at 0 and 15 dB: (a) average PAPR versus SNR and (b) conditional and mixture CCDFs.
	This configuration is independent of the main reliability lookup table in \figref{fig:throu}.}
	\label{fig:PAPR}
\end{figure}

\par
The PAPR-specific results are shown in \figref{fig:PAPR}.
Panel (a) exhibits step changes because this experiment selects among $\{1,0.95,0.9\}$ at 0 and 15 dB.
These modes and switching points are separate from the main reliability lookup table used in \figref{fig:throu}.
Panel (b) shows the fixed-mode conditional CCDFs and their occupancy-weighted mixture.
The curves characterize waveform peaks only; amplifier back-off and energy consumption require a specified nonlinear PA model and are not inferred here.

\section{Conclusion}
	\label{sec:conclusion}
		An SNR-aware FFTN-OTFS framework has been developed for high-mobility LEO satellite communications.
		A perception-reasoning-decision mechanism has been established using estimated SNR and offline BER thresholds to select the packing factor under a prescribed reliability requirement.
		Elevation-dependent propagation, fractional Doppler, FTN-induced ISI, and colored matched-filter noise have been incorporated into the signal model.
		Raw and effective throughput expressions and BER approximations for covariance-aware LMMSE detection have been derived, while EE and PAPR have been evaluated as supporting metrics.
		Numerical results have characterized the effects of SNR-estimation uncertainty on packing selection and channel-estimation error on detection.
		Under the tested channel configurations, the proposed strategy has restricted aggressive packing in unfavorable link conditions and has enabled denser signaling when the required SNR thresholds have been reached.
		These findings have highlighted the potential of reliability-constrained FTN adaptation for efficient data transmission in future LEO satellite networks.

\end{document}